\documentclass{elsart}
\usepackage{epsfig}

\begin{document}

\begin{frontmatter}
\title{A study of the proton spectra following the capture of
$K^-$ in  $^6$Li
and $^{12}$C with FINUDA}

\centering{FINUDA Collaboration}

\author[polito]{M.~Agnello},
\author[victoria]{G.~Beer},
\author[lnf]{L.~Benussi},
\author[lnf]{M.~Bertani},
\author[korea]{H.C.~Bhang},
\author[lnf]{S.~Bianco},
\author[unibs]{G.~Bonomi},
\author[unitos]{E.~Botta},
\author[units]{M.~Bregant},
\author[unitos]{T.~Bressani},
\author[unitos]{S.~Bufalino},
\author[unitog]{L.~Busso},
\author[infnto]{D.~Calvo},
\author[units]{P.~Camerini},
\author[infnto]{P.~Cerello},
\author[uniba]{B.~Dalena},
\author[unitos]{F.~De~Mori},
\author[uniba]{G.~D'Erasmo},
\author[uniba]{D.~Di~Santo},
\author[infnba]{D.~Elia},
\author[lnf]{F.~L.~Fabbri},
\author[unitog]{D.~Faso},
\author[infnto]{A.~Feliciello},
\author[infnto]{A.~Filippi\thanksref{corresponding}},
\author[infnpv]{V.~Filippini\thanksref{deceased}},
\author[infnba]{R.A.~Fini},
\author[uniba]{E.~M.~Fiore},
\author[tokyo]{H.~Fujioka},
\author[lnf]{P.~Gianotti},
\author[infnts]{N.~Grion},
\author[lnf]{O.~Hartmann},
\author[jinr]{A.~Krasnoperov},
\author[infnba]{V.~Lenti},
\author[lnf]{V.~Lucherini},
\author[infnba]{V.~Manzari},
\author[unitos]{S.~Marcello},
\author[kek]{T.~Maruta},
\author[teheran]{N.~Mirfakhrai},
\author[cnr]{O.~Morra},
\author[kek]{T.~Nagae},
\author[triumf]{A.~Olin},
\author[riken]{H.~Outa},
\author[lnf]{E.~Pace},
\author[lnf]{M.~Pallotta},
\author[uniba]{M.~Palomba},
\author[infnba]{A.~Pantaleo},
\author[infnpv]{A.~Panzarasa},
\author[infnba]{V.~Paticchio},
\author[units]{S.~Piano},
\author[lnf]{F.~Pompili},
\author[units]{R.~Rui},
\author[uniba]{G.~Simonetti},
\author[korea]{H.~So},
\author[jinr]{V.~Tereshchenko},
\author[lnf]{S.~Tomassini},
\author[kek]{A.~Toyoda},
\author[infnto]{R.~Wheadon},
\author[unibs]{A.~Zenoni}

\thanks[corresponding]{corresponding author. e-mail: filippi@to.infn.it; fax:
+39.011.6707324.}
\thanks[deceased]{deceased}
\address[polito]{Dip. di Fisica Politecnico di Torino, via Duca degli Abruzzi
Torino, Italy, and INFN Sez. di Torino, via P. Giuria 1 Torino, Italy}
\address[victoria]{University of Victoria, Finnerty Rd.,Victoria, Canada}
\address[lnf]{Laboratori Nazionali di Frascati dell'INFN, via E. Fermi 40
Frascati, Italy}
\address[korea]{Dep. of Physics,
Seoul National Univ., 151-742 Seoul, South Korea}
\address[unibs]{Dip. di Meccanica, Universit\`a di Brescia, via Valotti 9 Brescia, Italy and INFN Sez. di Pavia, via Bassi 6 Pavia, Italy}
\address[unitos]{Dipartimento di Fisica Sperimentale, Universit\`a di
Torino, via P. Giuria 1 Torino, Italy, and INFN Sez. di Torino,
via P. Giuria 1 Torino, Italy}
\address[units]{Dip. di Fisica Univ. di Trieste, via Valerio 2 Trieste, Italy and INFN, Sez. di Trieste, via Valerio 2 Trieste, Italy}
\address[unitog]{Dipartimento di Fisica Generale, Universit\`a di
Torino, via P. Giuria 1 Torino, Italy, and INFN Sez. di Torino,
via P. Giuria 1 Torino, Italy}
\address[infnto]{INFN Sez. di Torino, via P. Giuria 1 Torino, Italy}
\address[uniba]{Dip. InterAteneo di Fisica, via Amendola 173 Bari, Italy and INFN Sez. di Bari, via Amendola 173 Bari, Italy }
\address[infnba]{INFN Sez. di Bari, via Amendola 173 Bari, Italy }
\address[infnpv]{INFN Sez. di Pavia, via Bassi 6 Pavia, Italy}
\address[tokyo]{Dep. of Physics Univ. of Tokyo, Bunkyo Tokyo 113-0033, Japan}
\address[infnts]{INFN, Sez. di Trieste, via Valerio 2 Trieste, Italy}
\address[jinr]{JINR, Dubna, Moscow region, Russia}
\address[kek]{
High Energy Accelerator Research Organization (KEK), Tsukuba, Ibaraki
305-0801, Japan}
\address[teheran]{Dep of Physics Shahid Behesty Univ., 19834 Teheran, Iran}
\address[cnr]{INAF-IFSI Sez. di Torino, C.so Fiume, Torino, Italy
and INFN Sez. di Torino,
via P. Giuria 1 Torino, Italy}
\address[triumf]{TRIUMF, 4004 Wesbrook Mall, Vancouver BC V6T 2A3, Canada}
\address[riken]{RIKEN, Wako, Saitama 351-0198, Japan}

\begin{abstract}

Momenta spectra of protons emitted following the capture of $K^-$ in
$^6$Li and $^{12}$C have been measured with 1\% resolution. The $^{12}$C
spectrum is smooth whereas for $^6$Li a well defined peak appears at about 
500 MeV/$c$. The first 
observation of a structure in this region was identified as
a strange tribaryon or, possibly, a $\overline K$-nuclear state.
The peak is correlated with a $\pi^-$ coming from $\Sigma^-$ decay in flight, 
selected by setting momenta larger than 275 MeV/$c$. The $\Sigma^-$ could be
produced, together with a  500 MeV/$c$ proton, 
by the capture of a $K^-$ in a deuteron-cluster substructure of the 
$^6$Li nucleus.
The capture rate for such a reaction is 
$(1.62\pm 0.23_{stat}\; ^{+0.71}_{-0.44}(sys))\%/K^-_{stop}$,
in agreement with the existing observations on $^4$He targets and
with the hypothesis that the $^6$Li nucleus can be interpreted as
a $(d+\alpha)$ cluster.


\end{abstract}


\end{frontmatter}

\section{Introduction}
The capture process of a negative kaon in a nucleus
recently raised renewed interest in the search of the so-called
deeply bound $\overline{K}$-nuclear systems, predicted by Akaishi
and Yamazaki \cite{akaishi}. These states consist of few-body
strange (S=--1) systems composed of nucleons strongly bound to a
$\overline K$. The strength of the $\overline KN$ attractive
interaction, in the I=0 configuration, allows for the stability of
the system, as well as for its compactness \cite{dote}. The
binding energies predicted for these states are rather sizeable,
and the $\overline KN^{(I=0)}$ interaction should be so strong
(according to the hypothesis of these Authors) that the widths of
these aggregates may be very narrow. 
%In fact, the $\Lambda
%\pi$ decay channel, in whichever isospin configuration, is much
%weaker than the $\Sigma\pi$ one in $I=0$. However, 
In fact, the main decay mode of the $\overline K N$ system
is $\Sigma\pi$.
%, since the $\Lambda\pi$ channel in $I=0$ is much weaker. 
However, if the potential well is deep enough, even the 
$\Sigma\pi$ channel can turn out to be
energetically forbidden.
In this framework, these states are
expected to be formed with larger probabilities in lighter nuclei.

A signal with features compatible to these theoretical
expectations was recently found for the ($K^-$-three nucleon)
system by exploiting the missing mass method 
(KEK-PS E471 Experiment) \cite{ref2,ref3},  
and for the ($K^-$-two nucleon)
aggregate by studying the invariant mass of the system 
 (FINUDA at LNF) \cite{ref4}. 
$K^-$'s were stopped in both experiments, on a liquid
$^4$He target in the first case, and on thin solid targets of
several nuclear species ($^6$Li, $^7$Li, $^{12}$C) in the second. E471
claimed evidence for the existence of two different deeply bound
$K^-$-states, dubbed as $S^0(3115)$ and $S^+(3140)$, observed in
the inclusive proton and neutron spectra respectively. They
correspond to binding energies of about 193 and 169 MeV,
are less than 21 MeV wide and both decay  in the $\Sigma NN$
channel. On the other hand, FINUDA observed a structure decaying
into a back-to-back $\Lambda p$ pair at a mass of about 2255 MeV
(corresponding to a binding energy $B_{K^- pp} = 115
^{+6}_{-5}(stat)^{+3}_{-4}(sys)$ MeV), with $\Gamma =
67^{+14}_{-11}(stat)^{+2}_{-3}(sys)$ MeV.

A further hint for the existence of a strange multibaryon, namely
$^{15}_{K^-}$O, with a $K^-$
bound with a binding energy of $\sim 90\ (\mathrm{or}\ 120)$ MeV
was reported by Kishimoto {\it et al.} \cite{kishimoto} by exploiting
the $(K^-,\ n)$ reaction at 930 MeV/$c$, again based on a
missing-mass analysis.

Aside from the mentioned model, other
theoretical approaches exist.
Some of them admit the existence of bound kaon-nuclear states
but with shallower binding potentials \cite{re:shallow}; therefore
the expected widths of these states are too 
large to allow their experimental observation.

A dynamical approach, that allows the polarization of the nucleus
by the strong $K^-$-nucleus interaction, was recently proposed by
Mare$\check{\mathrm s}$ {\it et al.} \cite{galultimo}. In their
calculations the depth of the $\overline K$-nucleus potential is
density dependent, and the existence of rather
narrow (slightly less than 50 MeV wide) deeply bound kaonic states
is expected in nuclei heavier than $^{12}$C, for a deep binding
energy in the range $B_{\overline K} \sim (100-200)$ MeV. This
approach suggests that, for the much stronger nuclear polarization
obtained in Refs. \cite{ref2,ref3} for a target as light as
$^4$He, any $\overline K$-nucleus deeply bound state would be too
broad to allow clear observation.

Another theoretical argument against the existence of deeply bound
kaonic states was put forward by Oset {\it et al.}
\cite{osetprep}, who criticize the first theoretical approach
simply showing that the observed signals can be easily
explained in terms of few body $K^-$ absorption reactions
and their elementary
kinematics. The occurrence of $K^-$ absorption in nuclei with nucleons
emission is indeed a
well known process which occurs with a capture rate of the order of
10-20\%, so the possibility of observing pions and nucleons in the final
state of such processes is sizeable, and this simple
interpretation seems to be quite realistic.

We must also
recall that the experimental observations obtained so far are not
in perfect agreement with the Akaishi-Yamazaki model
expectations. In fact, E471 $S^0$ and $S^+$ states have a binding
energy much larger than the predictions from the model.
Moreover, their isospin is necessarily equal to one, 
contrary to the expectation of a singlet isospin
configuration to preferentially allow for deeply bound states.
Even the observation of a $(K^-pp)$ bound system by FINUDA is not
completely consistent with the features required by this model, especially
as far as the isospin and the expected width are concerned.
Therefore, the quest for the existence of such states is still open
and experimental confirmations are eagerly awaited. 

In this paper we report the results of a study of the proton
spectra (inclusive and semi-inclusive) emitted in the interaction
of $K^-$ at rest with $^6$Li and $^{12}$C nuclei, to be compared
to the results of Ref. \cite{ref2,ref3} on $^4$He.
The same missing mass
analysis technique followed by the Authors of Ref.
\cite{ref2,ref3} has been basically applied in the present case
also. First results of this analysis are reported in Ref.
\cite{re:hadron}.

\section{Experiment, data selection and spectrum shape analysis}

The experiment was performed with the  magnetic
spectrometer FINUDA, dedicated to Hypernuclear Physics and
installed at the DA$\Phi$NE $\phi$ factory at Laboratori Nazionali
di Frascati of INFN. Details on the structure of the FINUDA
spectrometer, on the measured performance as well as on the
conditions of the data taking in the 2003-2004 exploratory run 
can be found in Refs.
\cite{ref6,ref7,ref8}. We recall that the detector features 
momentum resolution from $0.5$ to $1.5 \%$ fwhm (depending on
the mass and momentum of the charged particle), an angular
acceptance close to $2\pi$ sr and  good particle
identification. The main apparatus features are briefly sketched
in the following. From the beam axis outward three parts can be 
distinguished, as described hereafter.

\begin{itemize}
\item {\it The interaction/target region}. It is located 
at the center of the apparatus.
Here the kaons from the $\phi$ decay are identified by means of a
12-thin-slab scintillator array, with the resolution $\sigma\sim
250$ ps. This barrel is surrounded by an octagonal array of
Silicon microstrips (ISIM, acronym for Internal SIlicon Microstrips) 
providing a first precise ($\sigma\sim 30$
$\mu$m) position measurement for the determination of the
interaction point of the $K^+,\; K^-$ pairs from of the $\phi$
decay, in the facing target modules. The target array consists
of eight slabs facing the microstrip modules and placed a few
millimeters from them, thin enough to stop the
$K^-$ as close as possible to the outer surface of the tile.
Different solid materials have been chosen as targets in the 2003-2004
data taking ($^6$Li, $^7$Li, $^{12}$C, $^{27}$Al, $^{51}$V);
\item {\it The external tracking device}. It consists of four layers of
position sensitive detectors, arranged in coaxial geometry and
immersed in a Helium atmosphere to minimize multiple scattering,
inside a 1.0 T solenoidal magnetic field. Facing the target tiles
a further array of ten Silicon microstrip modules is placed, with
basically the same features as the internal ones (OSIM, Outer SIlicon
Microstrips). The next
tracking devices are:
\begin{description} 
\item{i)} two octagonal arrays of He-iC$_4$H$_{10}$
(70-30) filled low mass drift chambers, featuring a spatial
resolution $\sigma_{\rho\phi}\sim 150$ $\mu$m and $\sigma_z\sim
1.$ cm; 
\item{ii)} a straw tube detector, at 1.1 m from the beam axis, 
composed of six layers of longitudinal and stereo tubes providing
a spatial resolution $\sigma_{\rho\phi}\sim 150$ $\mu$m and
$\sigma_z\sim 0.5$ cm;
\end{description}
\item {\it The external time-of-flight detector}. The outer TOF detector 
is a barrel of 72 scintillator
slabs, 10 cm thick, providing signals for the first level trigger
and time of flight measurements, with a time
resolution of $\sigma \sim 350$ ps. It allows also the detection
of neutrons with a $\sim 10\%$ efficiency and an energy resolution
of 8 MeV fwhm, for 80 MeV neutrons.
\end{itemize}

Out of five different nuclear species used for the eight targets, 
only $^6$Li and $^{12}$C were considered in the present analysis. Data
collected in the 2003-2004 exploratory run were used.

\begin{figure}[h]
\begin{center}
\begin{tabular}{cc}
\epsfig{file=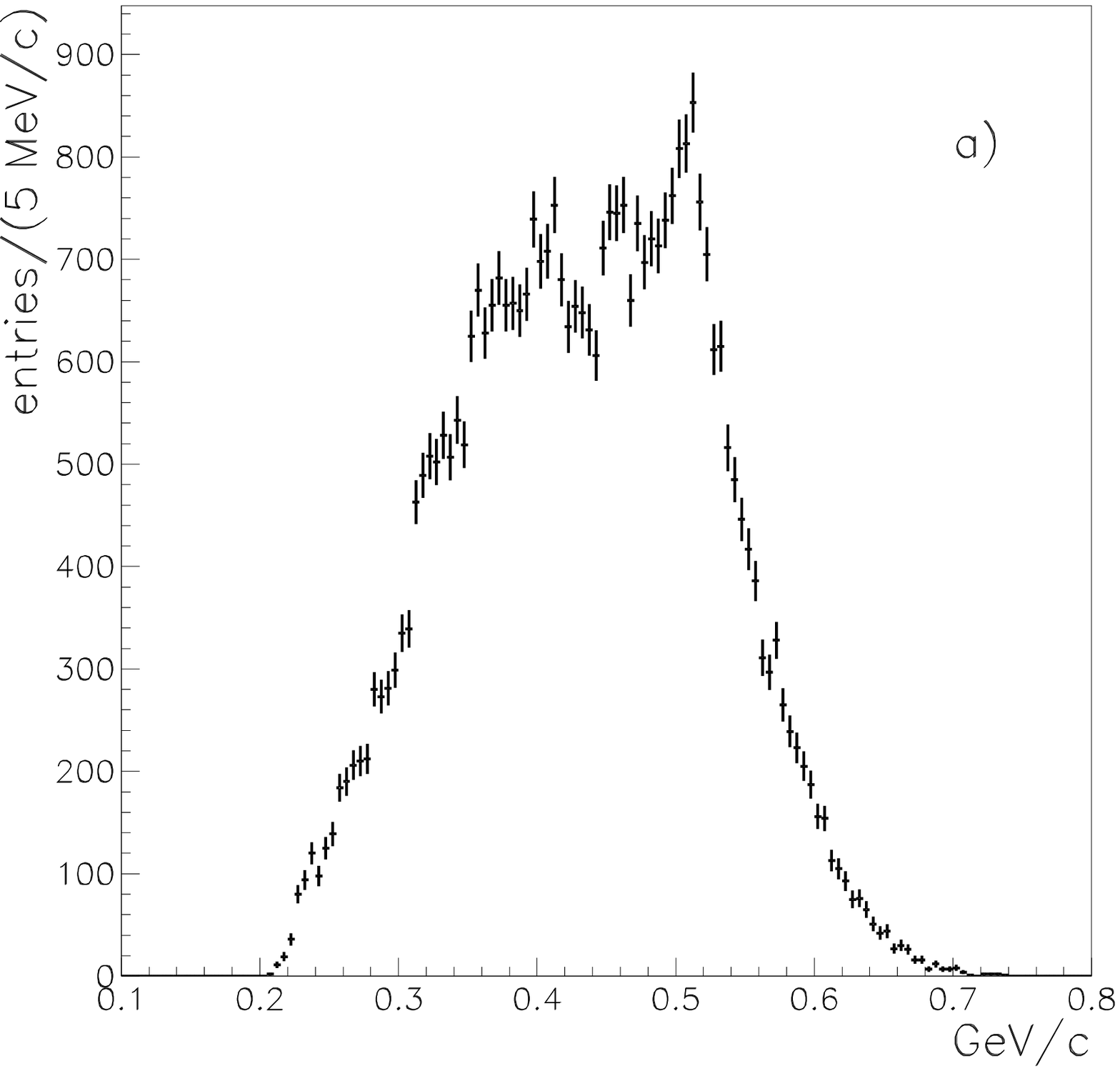,width=6truecm,height=5truecm,clip=}  &
\epsfig{file=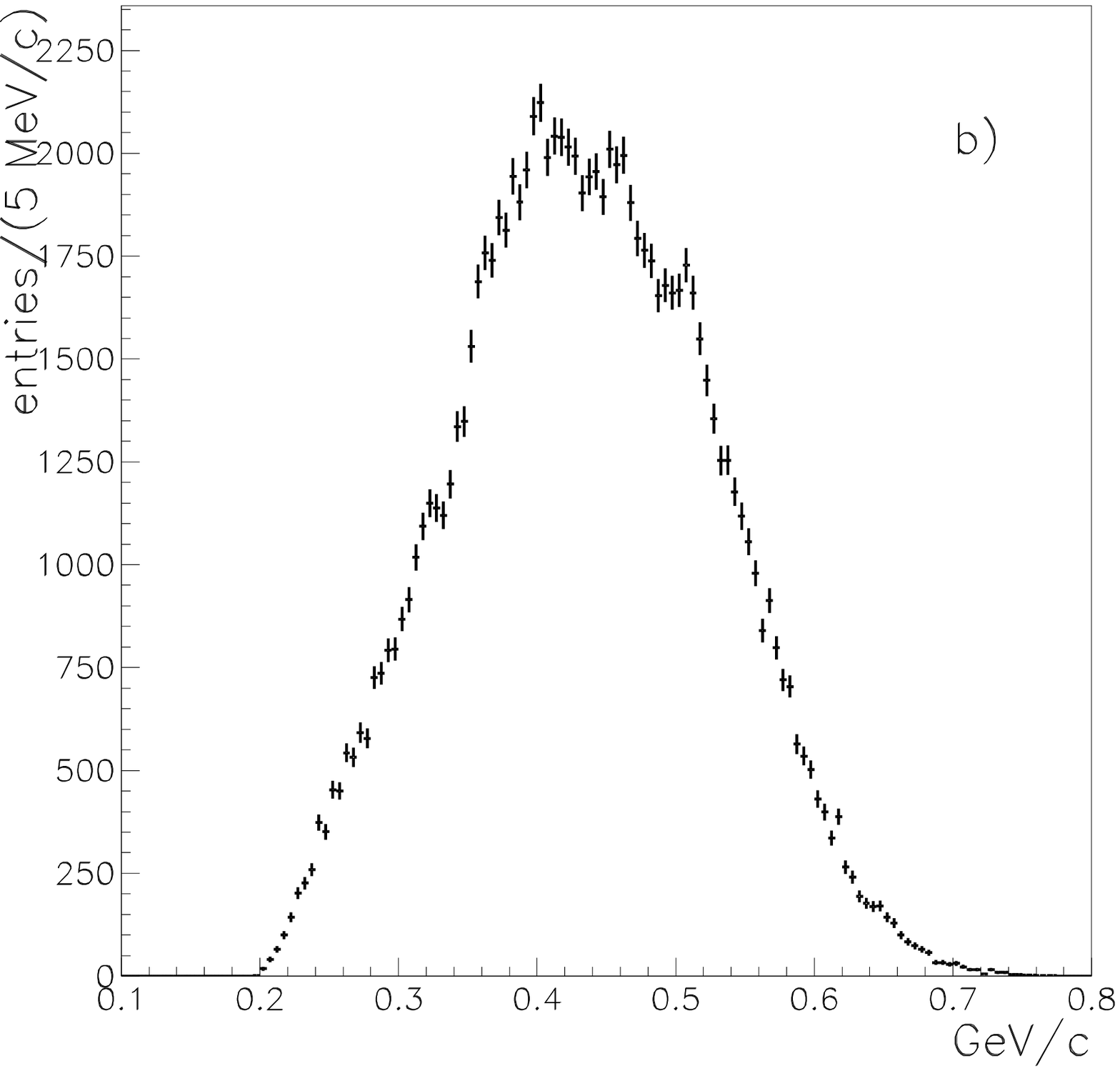,width=6truecm,height=5truecm,clip=} \\
\end{tabular}
\end{center}
\caption{Inclusive proton spectra measured
following the $K^-$ capture at rest from $^6$Li (a) and $^{12}$C
(b).} \label{fig1}
\end{figure}

Figure \ref{fig1}a) shows the inclusive proton spectra for $^6$Li
(two targets) and Fig. \ref{fig1}b) for $^{12}$C (three targets),
not acceptance corrected. The acceptance correction is effective
only in the lower momentum region of both plots, and doesn't
affect the shape in the region beyond 400 MeV/$c$, where it is
basically constant. In the lower momentum region it is a monotonic
rising function.

The spectra correspond to $(3307\pm 2_{stat}\pm 165_{sys})\times 10^3$ 
reconstructed $K^-$ stops  in two
$^6$Li targets, and $(4831\pm 2_{stat}\pm 142_{sys})\times 10^3$ 
ones in three $^{12}$C targets. A systematic error of 5\%
can be assigned to these numbers
due to a wrong $K^+/K^-$ identification, based only on kinematics and the 
position information provided by ISIM, and consequently a
wrong assignment of their stopping point in the targets.  

The events were selected applying the following criteria:
\begin{description}
\item{i)} high quality tracks, {\it i.e.}
long tracks crossing the full spectrometer with a hit on each of
the four tracking detectors, selected with the strict requirement
of a good $\chi^2$ out of the track fitting procedure;
\item{ii)} at least one proton per event recognized by particle identification
(energy loss in the outer Si-microstrip array)
with a probability $> 94\%$;
\item{iii)} a vertex reconstructed inside the fiducial volumes of the chosen targets.
\end{description}

The spectrum for $^{12}$C shows a smooth distribution with a
maximum around 400 MeV/$c$ and two distortions 
at about 450 and 520 MeV/$c$. The distribution is 
largely due to multinucleon $K^-$
absorption and, for a fraction, to non-mesonic decays of all kinds
of hyperfragments produced following the capture of $K^-$. 
A possible contribution due to protons originating from direct $(K^-, p)$
reactions in nuclei via weak interactions may be neglected due to the smallness
of the expected capture rate. The lower cut at $\sim
200$ MeV/$c$ is due to the momentum acceptance for four-hit
tracks, determined by the size of the spectrometer and the value
of the solenoidal magnetic field.

The spectrum of
$^6$Li shows, in addition to protons from multinucleon $K^-$ absorption
and from
non-mesonic decay of the $^5_{\Lambda}$He, $^4_{\Lambda}$He and
$^4_{\Lambda}$H hyperfragments (populating the region between
300 and 450 MeV/$c$), a prominent and narrow structure peaked at
$\sim 500$ MeV/$c$.

A feeble signal can also be distinguished at about 580 MeV/$c$.
Probably it is due to the absorption reaction
$K^-(NN)\rightarrow\Lambda N$, but it's also worth noting that in
this region one would expect the emission of deuterons in the rare
decay of the hyperhelium fragment
$^4_\Lambda{\mathrm{He}}\rightarrow d+d$ (570 MeV$/c$ being the
expected momentum value of the emitted deuteron). 
However, due to the smallness of the expected branching ratio, it 
is quite unlikely that this reaction could contribute to our inclusive
spectrum.

We recall that in Ref. \cite{ref4} a hint for the existence of
a $(K^-pp)$ bound state decaying into $\Lambda p$ was put forward. In this
case, due to the estimated binding energy of the state (about 110 MeV),
the proton momentum should be peaked at $\sim 420$ MeV/$c$. The protons from
its decay are of course included in Fig \ref{fig1}a) but, given a reported
capture rate of the order of $10^{-3}$/(stopped $K^-$), one can safely
ignore the contribution of such a channel.

\section{Data interpretation and discussion}

The peak around 500 MeV/$c$  in the $^6$Li spectrum is at about
the same position as the inclusive peak observed by E471 in Ref.
\cite{ref2}. Before giving physical interpretations to it, a
number of instrumental or elementary physical effects as the possible
origin of this signal were studied, and discarded for the reasons
discussed in the following.

A first suspect for a possible fake effect comes from the
observation that 510 MeV/$c$ is just the momentum of the $(e^+,
e^-)$ beams stored in the DA$\Phi$NE collider to produce the
$\phi(1020)$, whose decay into $(K^+, K^-)$ is the source of $K^-$
for FINUDA. Besides the argument that, if any positron leakage
could affect this analysis, one should observe such an effect in
both $^6$Li and $^{12}$C
 targets, a clear-cut consideration comes from the analysis of the
momentum spectrum of negative particles, selected with the same
criteria used to recognize the protons. No events were found
by requiring negative tracks having the same energy release in the
outer array of Si-microstrips as that required for protons. Due
to the symmetric behavior of $e^+$ and $e^-$, this possible source
of fake events can be safely discarded.

A physical phenomenon which could be a source of 508 MeV/$c$
protons is the ``rare'' two-body decay $^4_{\Lambda}{\mathrm{He}}
\rightarrow\ ^3{\mathrm H} + p$, that should appear in the
spectrum from $^6$Li but not in   $^{12}$C  since $^4_{\Lambda}$He
is one of the few hyperfragments produced from a $^6$Li target,
but  many others may be produced from $^{12}$C (the relative
production rates account for the different shapes of the spectra in
Fig. \ref{fig1}). This proton source was estimated as negligible
by the authors of Ref. \cite{ref2,ref3} by analyzing the light
outputs from the scintillators in order to select ``fast'' pions
accompanying the formation of $^4_{\Lambda}$He. With FINUDA, a
very powerful criterion could be adopted exploiting the information 
provided by the Si-microstrip modules. The energy release of the hits
in the ISIM moduli facing the $^6$Li targets emitting the proton 
were asked to be consistent to the passage of a $\sim 510$ MeV/$c$
$^3$H particle.

The number of events fulfilling the condition of a proton in the
$(480-530)$ MeV/$c$ range, with a back-to-back particle with an 
energy release compatible to a triton 
was $(58\pm 8_{stat})$. Out of these events, only two have an additional
$\pi^-$ in coincidence. At this stage of the analysis it is rather
hard to convert this number into an upper limit for the
$^4_{\Lambda}{\mathrm{He}} \rightarrow ^3\mathrm{H} + p$ two-body
rare decay due to the uncertainties in the determination of the
$^4_{\Lambda}$He production rate. However it is clear that the
relative rate of this reaction is negligible and it cannot
therefore be the source of an effect as sizeable as the one
observed in the data.

To have a better insight into the features of the observed proton
peak from $^6$Li, a few scatter plots have been studied in order
to point out possible correlations between the particles in the
peak and other particles of the same event tracked and identified
in the apparatus.

\begin{figure}[h]
\begin{center}
\begin{tabular}{cc}
\epsfig{file=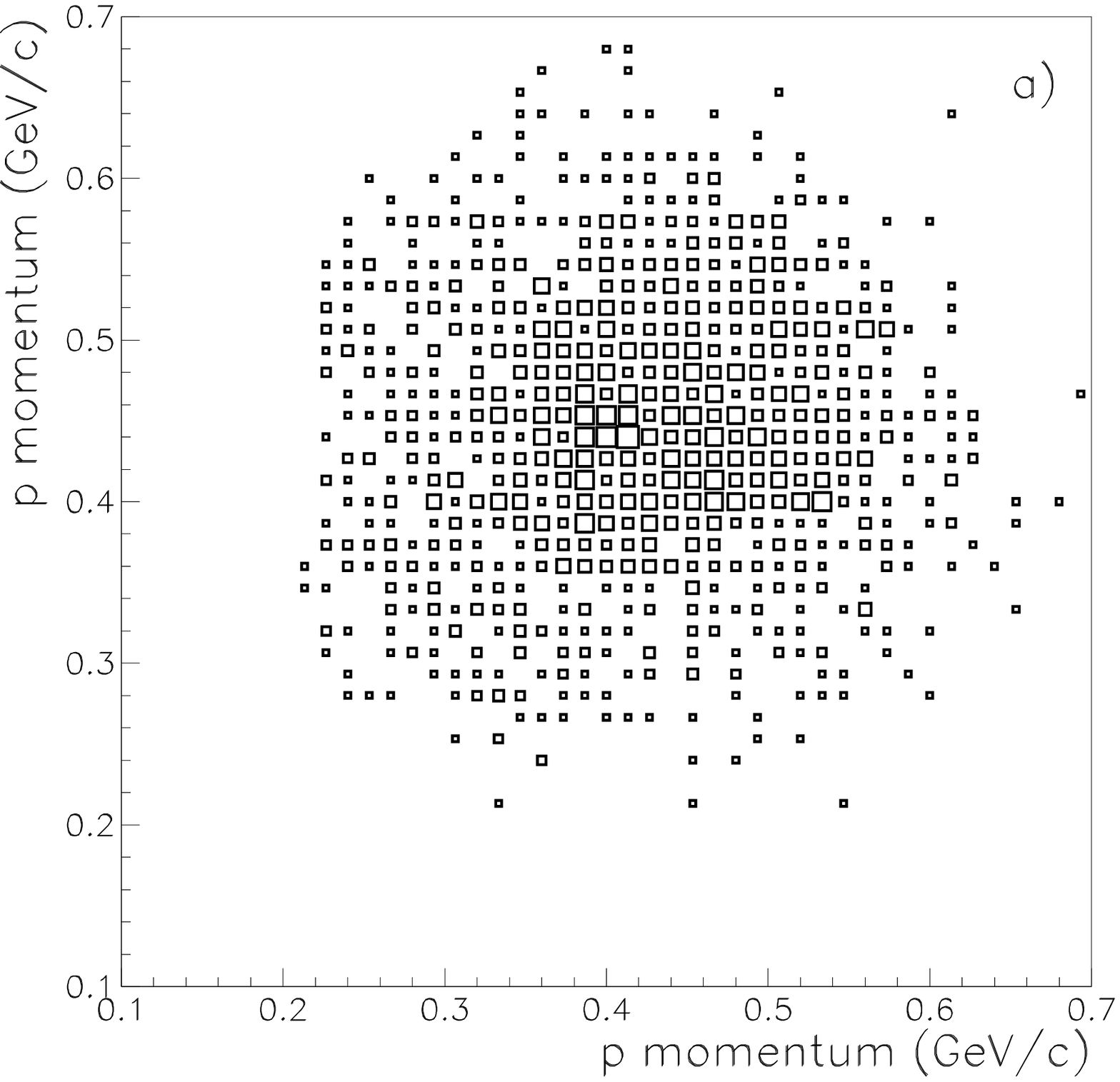,width=6truecm,height=5truecm,clip=}
&
\epsfig{file=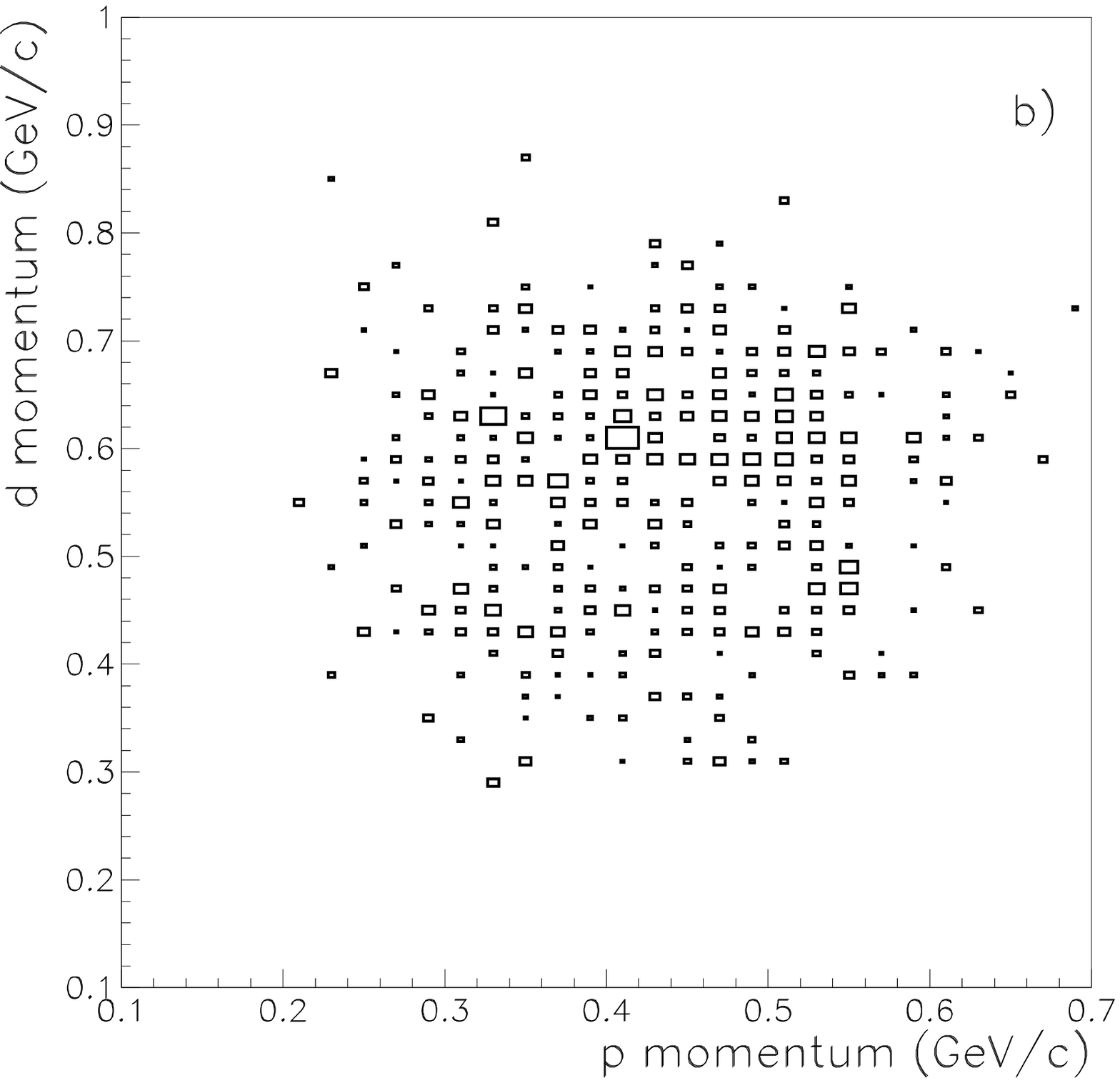,width=6truecm,height=5truecm,clip=} \\
\end{tabular}
\end{center}
\caption{Two-dimensional scatter plot of the proton momentum
versus a) a second proton track or b) a deuteron track, both
recognized by specific energy loss, and coming from a common
vertex in  $^6$Li targets.} \label{scatplots}
\end{figure}

The scatter plot of the two momenta for events with two tracks
recognized as two distinct protons is reported in Fig.
\ref{scatplots}a);  if the second track is identified as a
deuteron, the plot is shown in Fig. \ref{scatplots}b).

None of these shows any particular correlation. In Fig.
\ref{scatplots}a) two enhancements, almost symmetrical in the $x$ and
$y$ projections, can be seen in the $400-450$ MeV/$c$ range with
different intensities due to different reconstruction efficiencies
for the two tracks. These enhancements can be associated to the 
$K^-(pp)\rightarrow \Sigma^0 p$ reaction.
In Fig.
\ref{scatplots}b) the 500 MeV/$c$ proton signals are correlated to
a smooth phase space deuteron distribution, whose lower limit is
at about 300 MeV/$c$ and with a maximum at about 600 MeV/$c$. 

\begin{figure}[h]
\begin{center}
\begin{tabular}{cc}
\epsfig{file=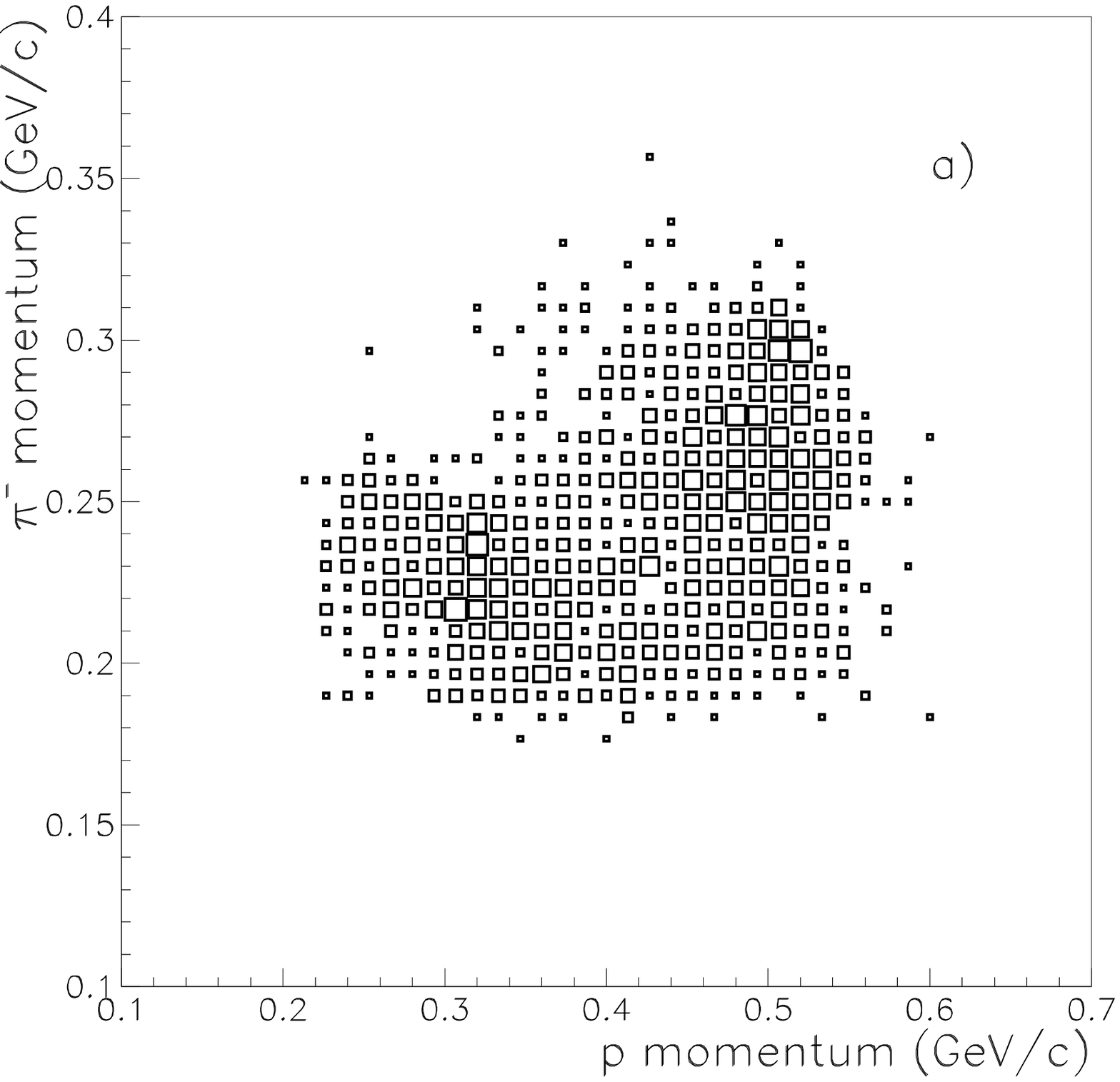,width=6truecm,height=5truecm,clip=}
&
\epsfig{file=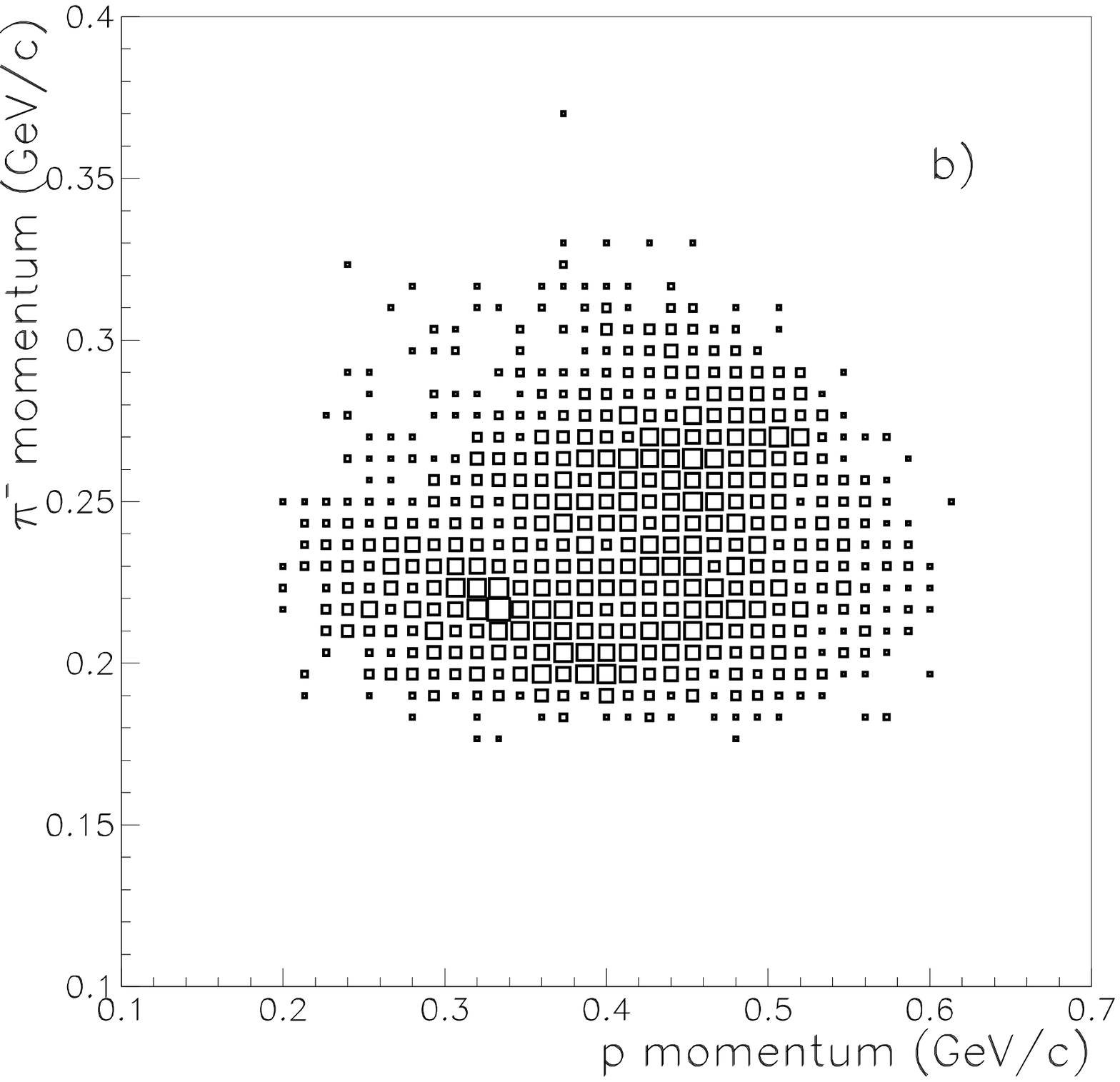,width=6truecm,height=5truecm,clip=} \\
\end{tabular}
\end{center}
\caption{Two-dimensional plot of the $\pi^-$ momentum versus the
proton momentum from $^6$Li (a) and $^{12}$C (b).} \label{fig2}
\end{figure}

More interesting is the correlation between the proton spectra of
Fig. \ref{fig1} and the momentum spectra of a $\pi^-$ emitted from
the same vertex. Fig. \ref{fig2}a) shows the scatter plot of the
$\pi^-$ versus proton momentum spectrum for $^6$Li, Fig.
\ref{fig2}b) for $^{12}$C. The main difference between the plots
is the presence of a correlation between high-momentum protons
$(\sim 500$ MeV/$c)$ with a high-momentum $\pi^-$ ($> 300$
MeV/$c$) in $^6$Li, that doesn't show up so clearly in $^{12}$C.

\begin{figure}[h]
\begin{center}
\begin{tabular}{cc}
\epsfig{file=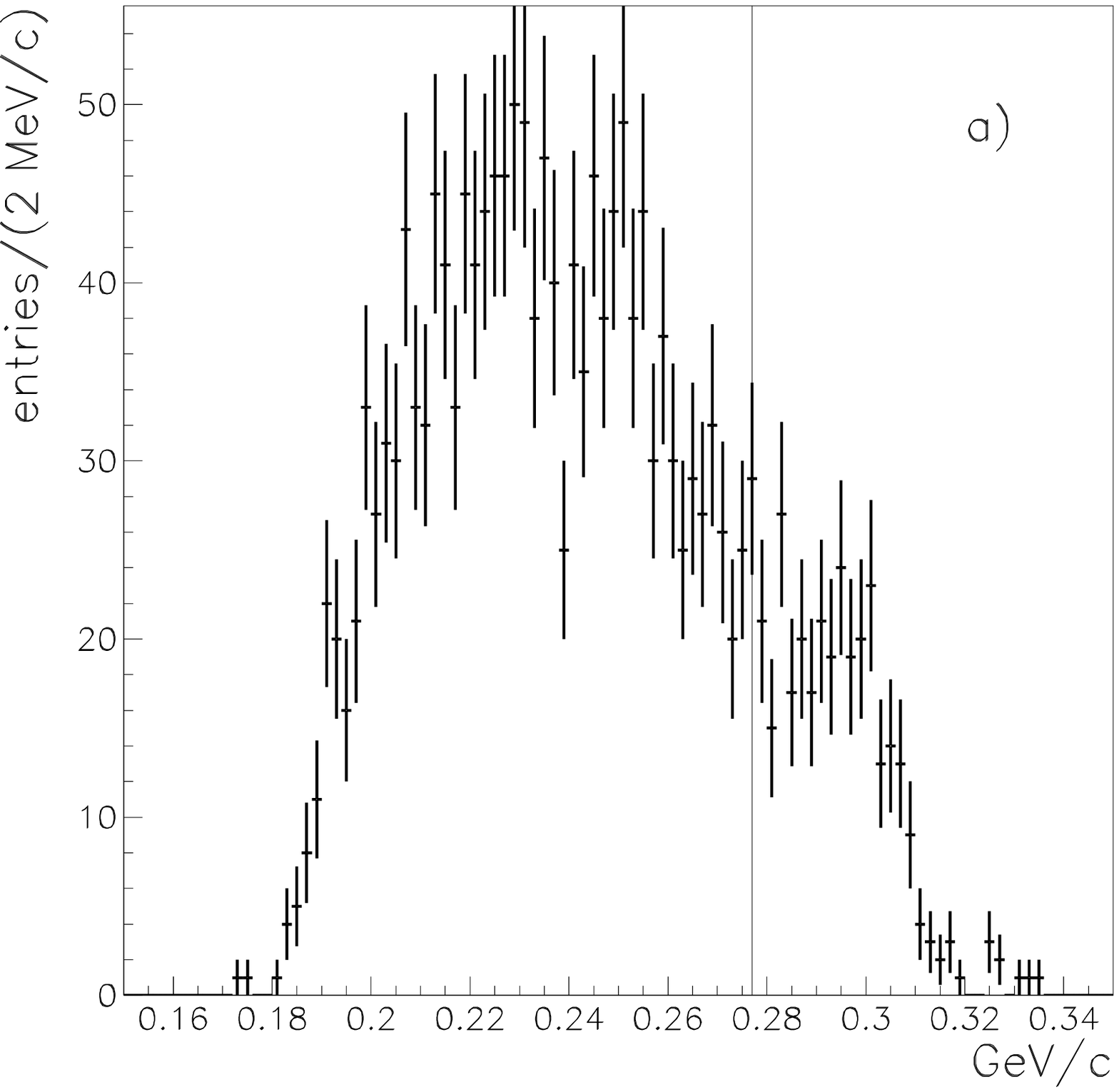,width=6truecm,height=5truecm,clip=}
&
\epsfig{file=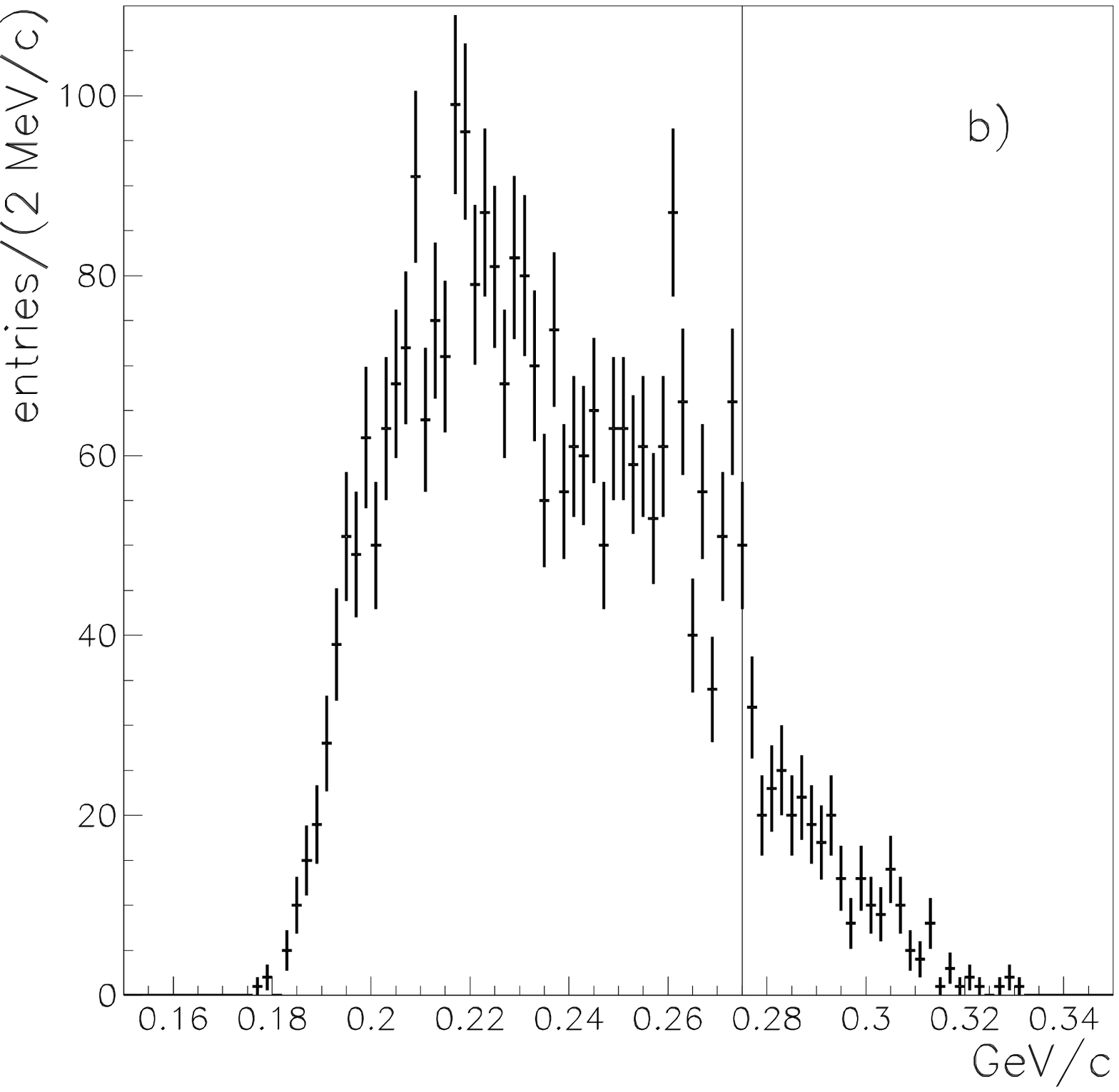,width=6truecm,height=5truecm,clip=} \\
\end{tabular}
\end{center}
\caption{Momentum spectrum of $\pi^-$ in coincidence with a
proton: $^6$Li (a) and $^{12}$C (b). The vertical lines indicate
the limit for $\Lambda$ production.  The narrow peaks in b) are
due to the formation of the  ground state and the 10 MeV
excited state of $^{12}_\Lambda$C.} \label{fig3}
\end{figure}

In figure \ref{fig3} we report the corresponding $\pi^-$ momentum
spectra for $^6$Li and $^{12}$C.

\begin{figure}[h]
\begin{center}
\begin{tabular}{cc}
\epsfig{file=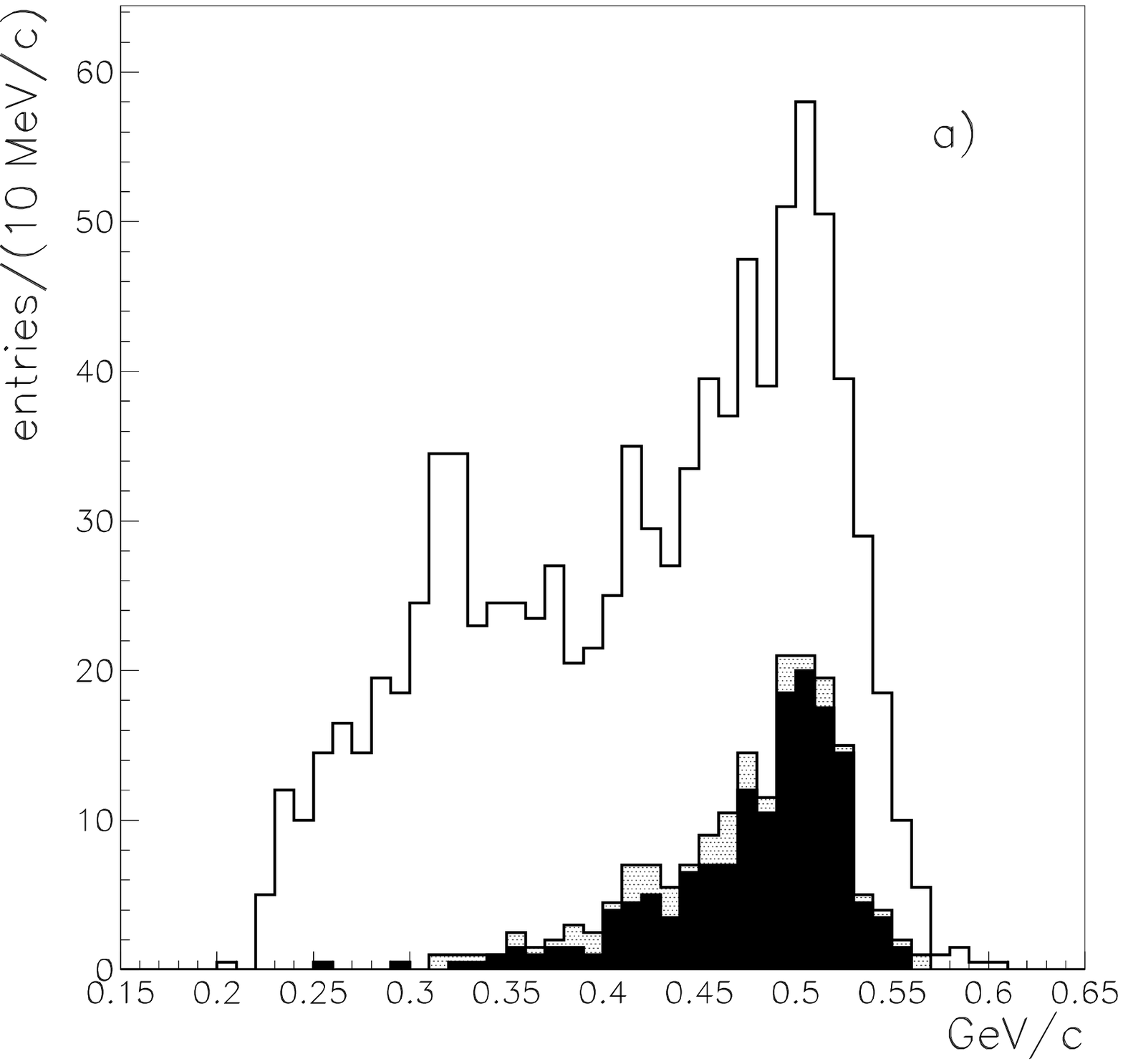,width=6truecm,height=5truecm,clip=}
&
\epsfig{file=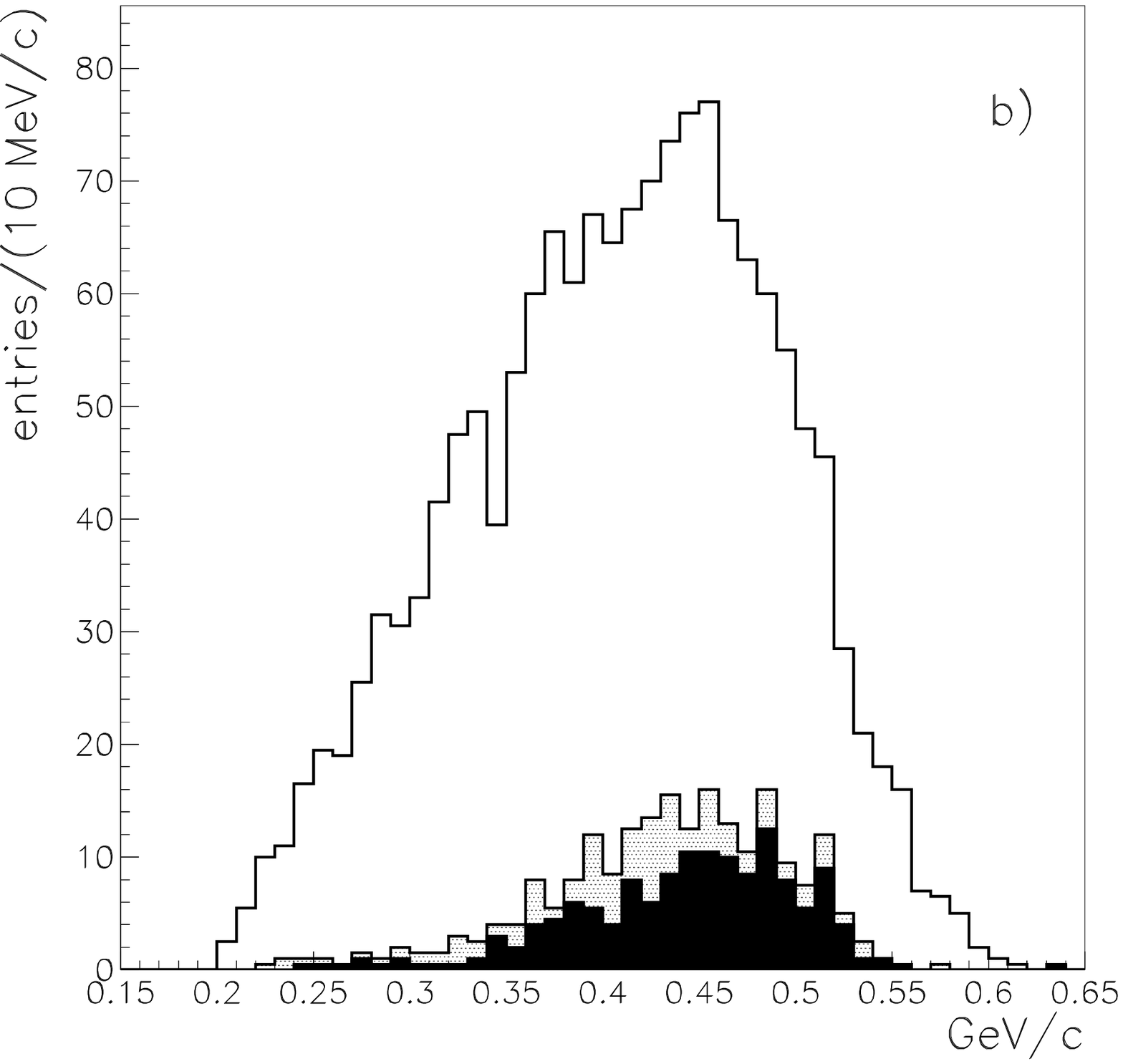,width=6truecm,height=5truecm,clip=} \\
\end{tabular}
\end{center}
\caption{Momentum spectrum of protons in coincidence with a
$\pi^-$: a) $^6$Li, b) $^{12}$C. The shaded spectra are obtained
for the condition of a $\pi^-$ momentum larger than 275 MeV/$c$
for $^6$Li and 272 MeV/$c$ for $^{12}$C. The black spectra are
obtained with the further condition of an angular correlation
$\cos\theta_{(\pi^- p)} < -0.8$.} \label{fig4}
\end{figure}

Figure \ref{fig4} shows (white histograms)  the proton momentum spectra
correlated with a $\pi^-$. The grey histograms represent the
proton momentum spectra when the coincidence $\pi^-$ has a
momentum larger than, respectively, 275 MeV/$c$ for $^6$Li and 272
MeV/$c$ for $^{12}$C (these values are marked in Fig. \ref{fig3}
with a vertical line).
This requirement is due to the fact that $\pi^-$'s with momenta
lower than 275 MeV/$c$ for $^6$Li and 272 MeV/$c$ for $^{12}$C
should be mostly associated with the production of
$\Lambda$-hyperfragments -- 275 MeV/$c$ is the kinematical limit for
the reaction $K^-_{stop} + ^6{\mathrm {Li}} \rightarrow
^5_{\Lambda}{\mathrm {He}} + p + \pi^-$, while  272 MeV/$c$ is the
limit for the reaction $K^-_{stop} + ^{12}{\mathrm C} \rightarrow
^{12}_{\Lambda}\mathrm{C}_{g.s.} + \pi^-$ \cite{ref7}.

A number of $K^-$ induced reactions on $^6$Li, or some
substructures of it ($^4$He, deuterons), have been studied in order
to identify the main source of the observed signature. Indeed, the
possibility of interaction of a $K^-$ on a substructure of a light
nucleus like $^6$Li is rather sizeable. In fact, in several
reactions a $^6$Li nucleus was observed to behave like a
$(d+\alpha)$ cluster, with the two nuclei in relative $s$-wave.
This behavior was in particular observed in $(\pi^+, 2p)$
reactions in flight \cite{ref14,ref15} and $(\pi^-, 2n)$ reactions
at rest \cite{ref16,ref17}. The same behavior holds also
for $^4$He, which can be  understood as a cluster of two
deuterons. The momentum distribution of a deuteron
cluster in $^6$Li has two peaks -- one at low momentum $(< 100$ MeV/$c$),
and the other at higher momentum ($\sim 300$ MeV/$c$), due to the 
``quasi''-deuteron contained in the $\alpha$ cluster 
(``quasi" here
indicates a few-nucleon cluster in a composite nucleus)
\cite{yamahiren}. On the other hand,
for the case of $^4$He, the low momentum component is missing and the deuteron
momentum distribution has a maximum at about 150 MeV/$c$. 

Therefore, the observed inclusive
proton spectra in $^6$Li could come from the interaction of a
$K^-$ with a nucleus as a whole, but also with one substructure of
it, a ``quasi''-$^4$He or a ``quasi''-deuteron. 
The first case could compare well to the 
$K^-$ $^4$He reaction (as studied from E471).
The different distribution of the deuteron
momentum in the two cases does not entail large distortions, except a
smearing of the signal observed in $^4$He as compared to the $^6$Li case,
so the structures observed in the inclusive spectra
out of $^6$Li and $^4$He could have the same nature. The possibility of
disentangling nuclear substructures stands only for light
nuclei. In nuclei as heavy as, for instance, $^{12}$C, were this
reaction still effective, the presence of the surrounding nucleons
would severely smear the signal making the observation of such a
reaction practically impossible. This fact agrees with the absence
of a clean signal in the momentum spectra of protons out of the
Carbon targets.

\begin{figure}[h]
\centering \epsfig{file=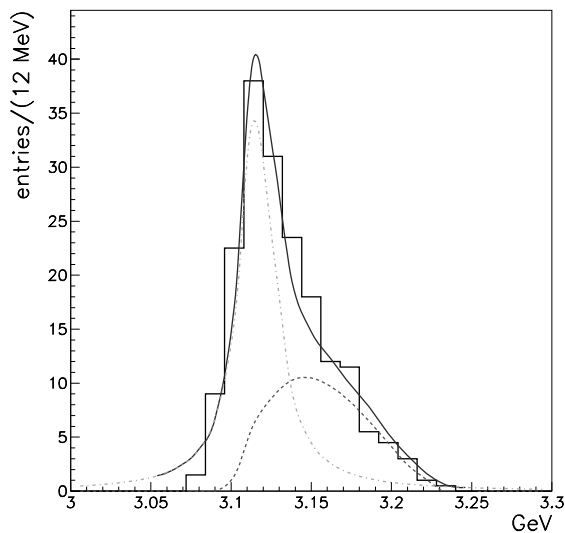,width=8truecm,clip=}
\caption{Missing mass spectrum obtained with the hypothesis of the
reaction $K^- + ^4{\mathrm{He}}\rightarrow X + p$, for data with
reconstructed vertices on $^6$Li targets; the $^4$He nucleus is
assumed to be at rest and not subject to any Fermi motion. The
dashed curve is a background shape fitting the analogous
spectrum obtained for $^{12}$C under the same hypothesis, adapted
to the tail of the $^6$Li distribution; the sum of this
contribution to a gaussian fitting the peak, properly weighted,
gives the solid line fitting the distribution.}
\label{missingMass}
\end{figure}

Using the FINUDA data, with the hypothesis that the grey spectrum
reported in Fig. \ref{fig4} is due to the $K^- + ^4{\mathrm{
He}}\rightarrow S^0 + p$ reaction, a fit with a Gaussian
function of the missing mass spectrum shown in Fig.
\ref{missingMass} delivers the values m=$(3116.4\pm 3.0)$ MeV and
$\sigma = (25.57\pm 0.54)$ MeV, respectively, as mass and width of
the hypothetical $S^0$ state (corresponding to a peak momentum
value of $(499.6\pm 2.2)$ MeV/$c$). These values were obtained
assuming that the signal stands on a background whose shape may be
roughly approximated by the shape of the analogous distribution
obtained from $^{12}$C data that is smoother, not peaked and
moved about 40 MeV forward compared to the one from $^6$Li. The
values obtained are in good agreement with those reported by E471
in Ref. \cite{ref2}. The contribution from the background amounts
to about 30\% of the integrated histogram.

 The $S^0$ is here
understood as a $(K^-pnn)$ bound state; its decay mode is assumed
to be the $YNN$ channel (for a discussion on the possible decay
for such states see \cite{dote,ref11}). The measured width is
larger than the upper limit of 21 MeV of Ref. \cite{ref2}, in
spite of the better momentum resolution. This effect could be
easily explained by referring to the  internal motion of the
``quasi"-$^4$He cluster inside $^6$Li.

\begin{figure}[h]
\centering \epsfig{file=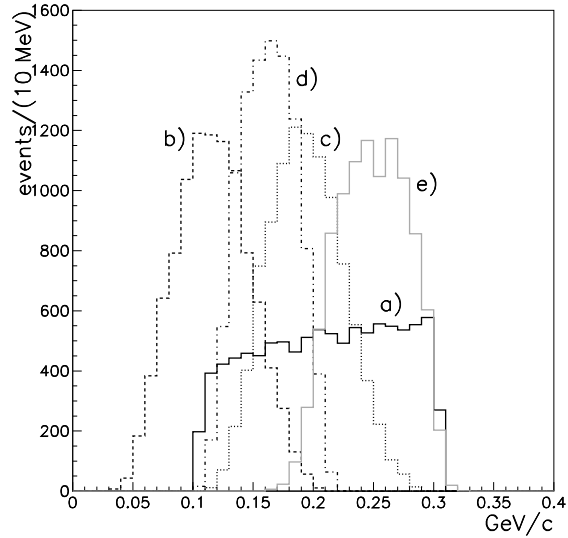,width=8truecm,clip=}
\caption{Monte Carlo momentum spectra for the $\pi^-$ coming from a hyperon
decay together with a $\sim 500$ MeV/$c$ proton in the following
reactions: a) black solid line: $K^- + d\rightarrow p + \Sigma^-$;
b) dashed line: $K^- + ^4{\mathrm {He}} \rightarrow S^0 + p$, with
the following $S^0\rightarrow \Lambda nn$ decay; c) dotted line:
$K^- + ^4{\mathrm {He}} \rightarrow S^0 + p$, with the following
$S^0\rightarrow \Sigma^- np$ decay; d) dotdashed line: $K^- +
^4{\mathrm {He}} \rightarrow S^0 + p$, with the following
$S^0\rightarrow \Sigma^- np$ decay and $\Sigma^- p\rightarrow
\Lambda n$ conversion; e) grey solid line: $K^- + ^4{\mathrm {He}}
\rightarrow S^0 + p$, with the following $S^0\rightarrow \Sigma^-
d$ decay. The spectra show the pure kinematics of the reaction
without any filtering through the apparatus acceptance and
reconstruction. The hyperon is always assumed to decay in flight.
The $^4$He and $d$ target nuclei are assumed to be at rest and not
embedded inside a composite nucleus. When they are endowed with a 
Fermi momentum, which
depends on the target nucleus, the distributions are moved toward
higher momenta  and are more smeared.} \label{spettriMC}
\end{figure}

The values found for its
mass and width were inserted in the simulations of the
kinematics of several reactions
aiming to single out the one that gives a
signature compatible with the observed distributions.
In Fig. \ref{spettriMC} a summary of the shape of the expected
$\pi^-$ spectra in the following $K^-$ induced reactions is
reported (the branching ratios are not normalized):
\begin{description}
\item{a)} $K^- + d\rightarrow p + \Sigma^-$, on a deuteron at rest:
the expected momentum for the
proton is 515 MeV/$c$, and the $\Sigma^-$ is forced to decay in
flight into $n+\pi^-$;
\item{b)} $K^- + ^4{\mathrm {He}} \rightarrow S^0 + p$, with the
following $S^0\rightarrow \Lambda nn$ decay;
\item{c)} $K^- + ^4{\mathrm {He}} \rightarrow S^0 + p$, with the
following $S^0\rightarrow \Sigma^- np$ decay;
\item{d)} $K^- + ^4{\mathrm {He}} \rightarrow S^0 + p$, with the
following $S^0\rightarrow \Sigma^- np$ decay and $\Sigma^-
p\rightarrow \Lambda n$ conversion;
\item{e)} $K^- + ^4{\mathrm {He}} \rightarrow S^0 + p$, with the
following $S^0\rightarrow \Sigma^- d$ decay.
\end{description}

The expected momentum of the proton recoiling against the
simulated $S^0$ is 501 MeV/$c$. The only reactions which could be
responsible for the observed signature are {\it a)}, {\it c)} and
{\it e)} -- that means that $\pi^-$ of momenta larger than the
above limits are associated to the in-flight decay of $\Sigma^-
\rightarrow n + \pi^-$ produced either in the quasi two-body reaction
$K^- + (np) \rightarrow \Sigma^- + p$ on correlated $(np)$ pairs,
or deuterons, in the nuclei, or following the decay of the deeply
bound $K^-$- nuclear states into $\Sigma^- NN$. All the nuclei are
assumed to be at rest, and no Fermi momentum is assigned in the
simulation to any of them.

An experimental confirmation that the $\pi^-$ comes from a hyperon decay
 and not exactly from the vertex in which the proton is emitted can be
obtained from the study of the impact parameter for the two
particles, emitted from the same $^6$Li target. 
The impact parameter is here defined as the distance of closest approach
between the $K^-$ stop point (calculated by tracking the particle
inside the target material with GEANE routines \cite{geane}) 
and the particle
track, extrapolated backwards to the target region, by means of
GEANE routines as well. 

\begin{figure}[h]
\begin{center}
\begin{tabular}{cc}
\epsfig{file=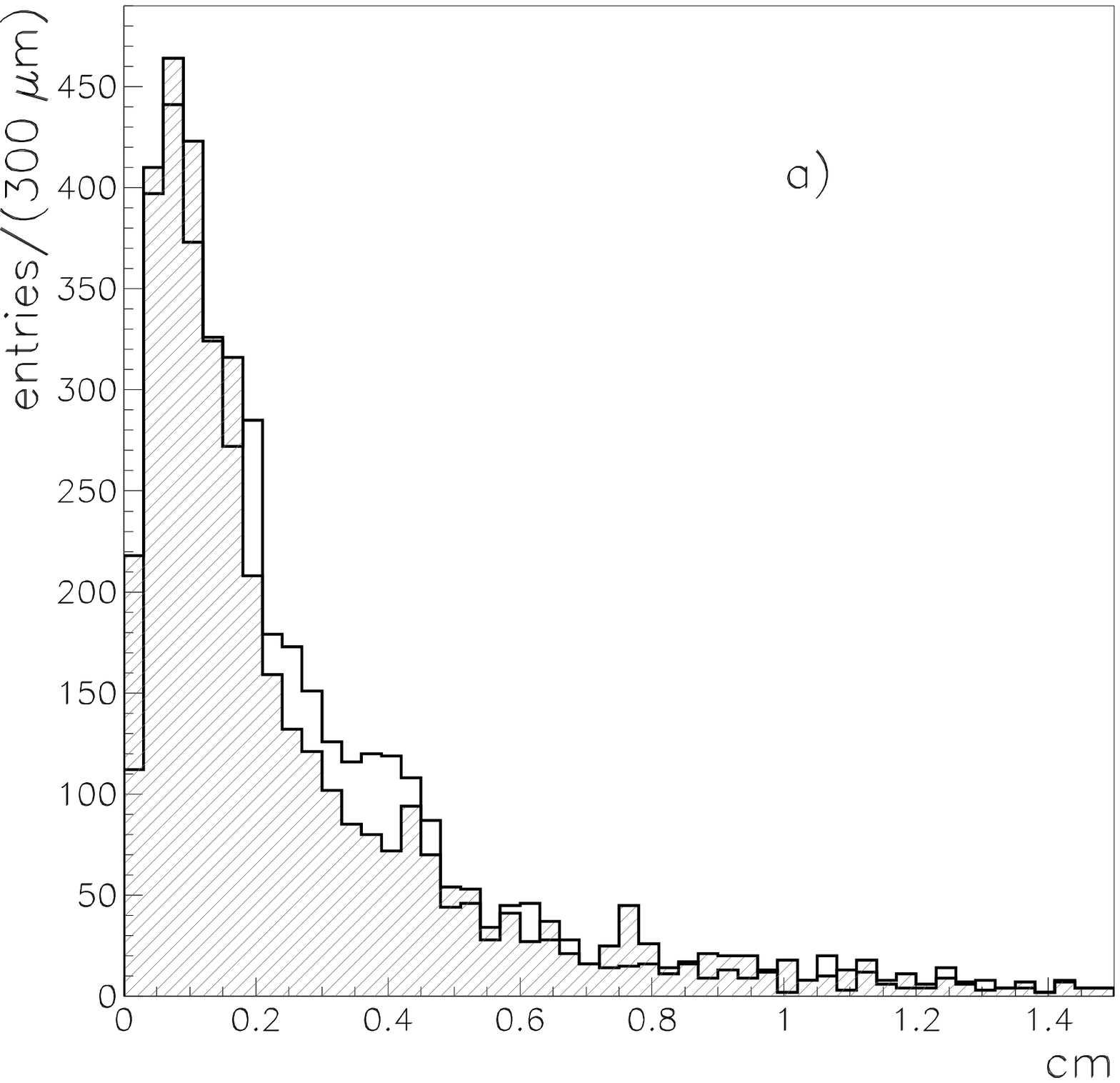,width=6truecm,height=5truecm,clip=}
&
\epsfig{file=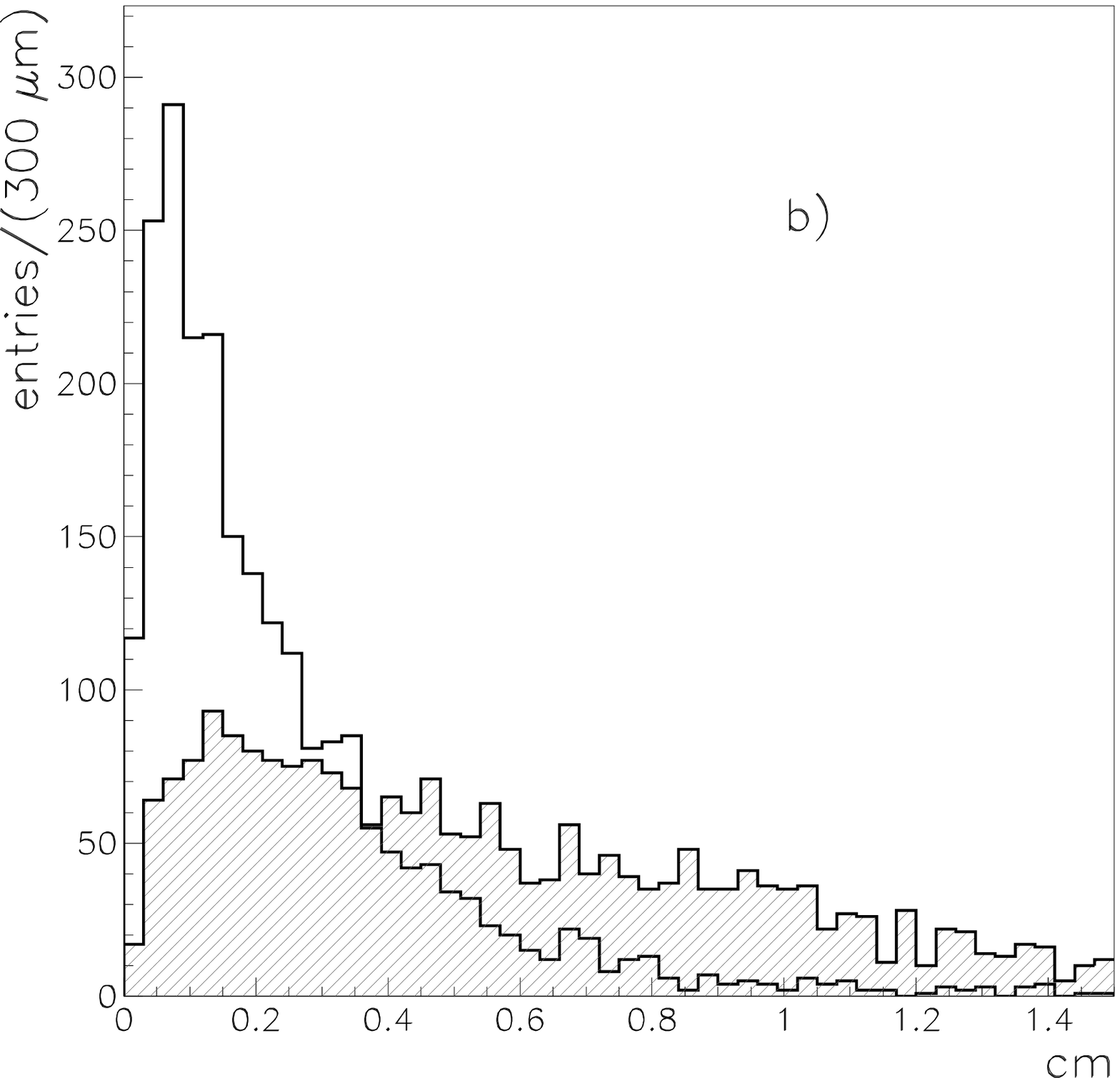,width=6truecm,height=5truecm,clip=} \\
\end{tabular}
\end{center}
\caption{Distributions of the impact parameter for Monte Carlo data,
after reconstruction, in the two cases: a) production of an
uncorrelated $\pi^-$-p pair from the $K^-$ vertex, and b) production
of $\Sigma^-$ and $p$ from the decay vertex, and following $\Sigma^-$
decay. The open histograms correspond to the proton, and the hatched ones
to the $\pi^-$ case.} \label{impParMC}
\end{figure}

Fig. \ref{impParMC} shows the expected distributions after the 
reconstruction in the FINUDA apparatus of Monte Carlo events with 
a) an incorrelated pion and proton pair
emitted from the same $K^-$ vertex, and b) 
a proton and a $\Sigma^-$ emitted from the $K^-$ vertex, 
and a $\pi^-$ coming from the $\Sigma^-$
decay. 

While in the first case no difference is seen in the 
distributions, in the second a substantial shape difference can be appreciated,
even if a direct estimation of the decay length cannot easily be deduced. 
It is noteworthy  how in the first case no sensible distortion
due to multiple scattering effects affect the two distributions.

\begin{figure}[h]
\centering \epsfig{file=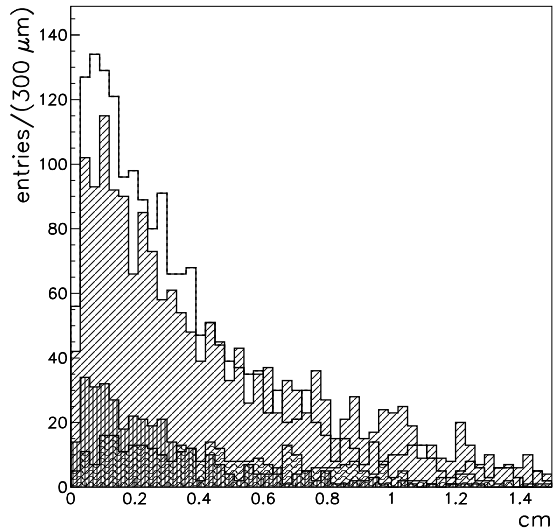,width=8truecm,clip=}
\caption{Distribution of the impact parameter for protons (white
histogram) and pions (hatched  histogram) following the interaction
of a $K^-$ in the $^6$Li targets. The vertical hatched histogram and
the wavy one correspond to the proton and $\pi^-$ impact parameter
distributions for events with $p_{\pi^-}>275$ MeV/$c$.} \label{impact}
\end{figure}

The distributions of the 
distance of closest approach from experimental data
are shown in Fig. \ref{impact}. The
shapes of the distributions for the proton (open histogram) 
and the $\pi^-$ (hatched histogram) are
different and suggest that the pion is most probably produced away
from the $K^-$ vertex, even if the reconstruction algorithm
associates the track to the same point.

In the same Figure, the vertical (wavy) hatched histogram shows the 
experimental distribution for the proton ($\pi^-$) momentum in events
where the $\pi^-$ momentum is required to be larger that 275 MeV/$c$,
corresponding to the shaded histogram in Fig. \ref{fig4}a). 
%In this case,
%a large difference between the two shapes can be observed, indicating
%that there is a good chance that the two particles are really emitted from
In this case,
a large difference between the two shapes can be observed, indicating
that there is a high probability that the two particles are really 
emitted from different points.

A handle to understand which of the reactions mentioned above could be
the most favorable source for the observed signature is given by
the study of the angular correlation (if any) between the $\pi^-$
and the proton. Were they coming from a reaction like {\it d)} or
{\it e)}, where the proton is emitted together with the deeply
bound $S^0$ state and the $\pi^-$ comes from the decay of a
$\Sigma^-$ that is one of the products of the decay at rest of the
$S^0$, no particular angular correlation is expected. Even in the
case of the formation of a more complicated deeply bound structure
in $^6$Li, made of five baryons, as could be obtained in the
reaction $K^- + ^6{\mathrm {Li}}\rightarrow (K^-ppnnn)+p$, no
angular correlation is expected between the recoiling proton and
the $\pi^-$ presumably coming from the decay of a hyperon among
the deeply bound kaonic state decay products.

On the other hand, for a two-body process like {\it a)} the proton
and the $\Sigma^-$ should be emitted back-to-back, and, if the
$\Sigma^-$ decays in flight, the $\pi^-$ should keep track of the
direction of travel of its mother, therefore an angular
correlation should be evident.

\begin{figure}[h]
\centering
\epsfig{file=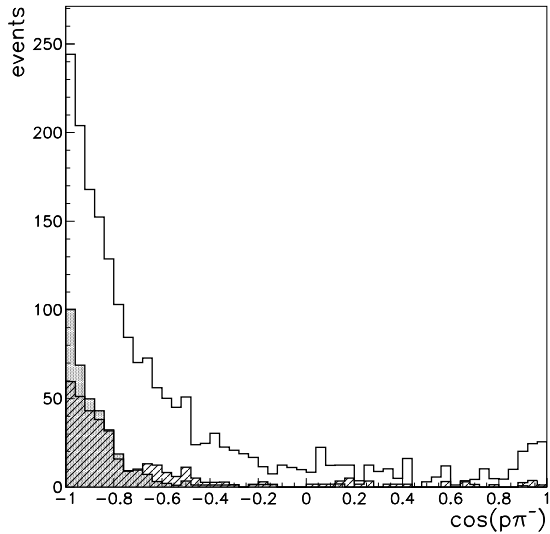,width=8truecm,clip=}
\caption{Distribution of the angle between a proton and a $\pi^-$ track
following the interaction of a $K^-$ in the $^6$Li targets: the white
histogram is for all events, the shaded one for events with the pion
momentum larger than 275 MeV/$c$, the hatched one for events with the proton
momentum in the window $(275-350)$ MeV/$c$.
All the distributions are acceptance corrected and normalized to the
number of selected events in each sample.}
\label{fig5}
\end{figure}

Such a correlation was searched for in the coincidence events giving
the spectra of Fig. \ref{fig4}a). In Figure \ref{fig5} the angle
between the proton and the $\pi^-$ track out of the $^6$Li targets
is shown, properly acceptance corrected. The white histogram
corresponds to all the events, while the shaded one to events with
a fast $\pi^-$ ($p_{\pi^-} > 275$ MeV/$c$). The effect of this cut
is to eliminate small angle events, keeping mainly back-to-back
correlated ones. By requiring the condition $\cos\theta_{(\pi^- p)} <
-0.8$, the black distributions shown in Fig. \ref{fig4} were
obtained for both $^6$Li and $^{12}$C targets: the peak in $^6$Li
data is still prominent and possibly shrinks, while this selection
has no particular effect on the distribution out of $^{12}$C.

This observation is the most convincing proof in favor of the
two-body reaction on two nucleons with  $\Sigma^- p$ production, and
against the production of a deeply bound $K^-$ state with  the
features suggested by E471. Since the missing mass spectrum
analysis pointed out a close familiarity between the signal
observed by FINUDA and the one seen by E471, it might be
plausible, in this view, that the peak observed in the $K^-$
$^4$He reaction could be due to the interaction of the $K^-$ on a
deuteron-like substructure of the E471 $^4$He target; therefore it
appears to be unnecessary to invoke the existence of one or even
two deeply bound $K^-$ states to explain it.

\subsection{Comments on the specrtum shape at lower momenta 
($\sim 300$ MeV/$c$).}

Figure \ref{fig4}a) shows clearly a further structure at about 300
MeV/$c$, that can  be interpreted as a further hint to a 
$(\alpha+d)$ behavior of $^6$Li in the $K^-$ capture. The
possibility that it might be due to an acceptance effect can be
rather safely discarded since, as such, it should also be visible
in the $^{12}$C case. Moreover, the acceptance function is this
region is steeply rising, but is smooth and doesn't exhibit any
structure.

Therefore, we suggest attributing it to the reaction $K^-+n\rightarrow
\Lambda+\pi^-$ occurring on a neutron of the deuteron cluster. The
internal momentum of such  a neutron in the quasi-deuteron
substructure is low, given {\it e.g.} by the well-known Hulth\'en
formula; the momentum of the quasi-deuteron relative to
the $\alpha$-cluster is very small, as previously mentioned \cite{ref15}. 
The total
momentum of the neutron inside a quasi-deuteron cluster
is lower than 100 MeV/$c$. So the
reaction on a neutron in a quasi-deuteron cluster shows basically 
the same kinematic features
of the reaction on a free neutron, 
with the momenta of both $\Lambda$ and $\pi^-$
around 250 MeV/$c$. The $\Lambda$ decays in flight in a
(forward emitted) proton with a momentum of about 300 MeV/$c$, and a
$\pi^-$ with a momentum around 110 MeV/$c$. This pion escapes the
momentum acceptance of the reconstruction procedure, and therefore goes 
undetected. In Fig.
\ref{fig3}a) a hint of a peak for $\pi^-$ around 250 MeV/$c$ can
be seen, and the scatter plot of Fig. \ref{fig2}a) also shows a
correlation between 300 MeV/$c$ protons and $\sim 250$ MeV/$c$
$\pi^-$.

The distribution of the angle between a proton with momentum
selected in the window $(275-350)$ MeV/$c$ and the $\pi^-$ is
reported in the hatched histogram in Fig. \ref{fig5}. Even in this
case there is a marked preference for a back-to-back topology,
consistent with the hypothesis that the proton comes from the
$\Lambda$ decay, with the $\Lambda$ emitted in the opposite
direction to the $\pi^-$, in the $K^-+n\rightarrow \Lambda+\pi^-$
reaction occurring almost at rest.

\subsection{Evaluation of the capture rate for the
$K^- + (np)\rightarrow \Sigma^- + p$ reaction in $^6$Li}

The evaluation of the capture rate for negative kaons inducing the
two-body reaction under examination is based on their tagging by
means of the $\mu^+$ or $\pi^+$ from the decay of the $K^+$
emitted in the $\phi$ decay. Out of the selected events, the
total number of kaons to which the selected sample is normalized is 
obtained selecting those having also a positive track correctly identified 
as a $\mu^+$ or $\pi^+$. This track must come from a target
opposite to the one where the $K^-$ stopping occurs.

The capture rate for the selected events in coincidence with a
$\mu^+$ or a $\pi^+$
from the opposite target ({\it tagged} events) is given by
\begin{equation}
R = \frac{N^{tag}_{ev}}{N(K^{-\; tag}_{stop})\cdot \epsilon_D(ev)\cdot
\epsilon^{tag}_G(ev)}
\label{eq:rcalc}
\end{equation}
where $N(K^{-\; tag}_{stop})$ is the number of $K^-$'s stopping in
the given target (for $\mu^+/\pi^+$ tagged events),
$\epsilon_D(ev)$ is the detector efficiency for the selected
events and $\epsilon_G(ev)$ is the global efficiency taking into
account the trigger and the reconstruction efficiency and the
geometrical acceptance of the apparatus. For events with a
coincidence $\mu^+/\pi^+$ the trigger efficiency is practically
equal to one, so $R$ can be considered, in this case, as
independent of the trigger efficiency. 
We recall that the trigger applied (mainly to select 
hypernuclear formation events) required two back-to-back slabs firing on
the internal scintillator barrel, 
with signal amplitude above an energy threshold 
accounting for the high ionization of slow kaons, and a fast coincidence on
the external scintillator barrel, 
to account for the presence of a $\mu^+$ from the $K^+$
decay.

The global efficiency for the selected events is given, as usual, by
\begin{equation}
\epsilon^{tag}_G(ev) = \frac{N_{ev}^{tag, MC}}{N(K^{-\; tag, MC}_{stop})},
\end{equation}

while the detector efficiency $\epsilon_D(ev)$ must be carefully
evaluated since the typical event is composed of two four-point
tracks almost back-to-back. For a four point track fully
reconstructed in a selected apparatus region (depending on the
target position), the detector efficiency can be deduced by the
simulation of well known reactions, such as  $K_{\mu 2}$ decay,
comparing the simulated and experimental decay branching ratios
obtained for well identified muons emitted from the same target.
One can roughly assume that the detector efficiency for the two
prongs of a typical event of the reaction under study may be given
by the combination of the efficiencies of two single tracks -- one
emitted in the forward direction, and one in the backward
hemisphere with respect to the chosen target. On an event-by-event
basis, it would be possible to assign to each single track the
correct forward/backward detector efficiency. In the rate
calculation, however, this assignment is not possible. So, as a first
approximation and on average, the detector efficiency for each
track is assumed to be equal to the efficiency for positive  muons
emitted in both the hemispheres, and crossing almost the
whole apparatus. However, the systematic error affecting this
approximate procedure in the detector efficiency evaluation
does not exceed  0.7\% of the evaluated rate.

In Tab. \ref{tab:numeri1}
the relevant numbers to get the $R$ values
for the two $^6$Li targets are reported.

\begin{table}[ht]
\centering
\begin{tabular}{lccc|c}
\hline
$N^{tag}_{ev}$ & $N(K^-_{stop, tag})$ & $\epsilon^{tag}_G(ev)\times 10^{3}$ &
$\epsilon_D(ev)$ &
$R/K^-_{stop}(\%)$ \\ \hline
$\scriptstyle 10\pm 3$ & 
$\scriptstyle 347174\pm 589_{stat} \pm 17359_{sys}$ 
& $\scriptstyle 9.0\pm 0.8$ & 
$\scriptstyle 0.229\pm 0.008$ 
& $\scriptstyle 1.43 \pm 0.62_{stat}\;^{+0.20}_{-0.07}(sys)$ \\
$\scriptstyle 23\pm 5$ & $\scriptstyle 363745\pm 603_{sys}\pm 18187_{sys}$ 
& $\scriptstyle 11.1\pm 0.8$ & 
$\scriptstyle 0.313\pm 0.008$ &
$\scriptstyle 1.85 \pm 0.56_{stat}\; ^{+0.30}_{-0.09}(sys)$ \\
\hline
\end{tabular}
\caption{Measured number of events, $K^-$ stops, efficiency
and capture rate for the reaction $K^- + (pn)
\rightarrow \Sigma^- p$. The number of events and of stops are
both tagged with a coincidence $\mu^+$ or $\pi^+$ on the opposite target.
The number of selected events for the reaction under study is already
background subtracted (the background amounts to about 30\% of the
observed signal).
The two lines correspond to events collected on two different $^6$Li targets.}
\label{tab:numeri1}
\end{table}

A second way to determine the capture rate of the studied reaction
is based on the normalization on the number of $\mu^+$'s produced
by $K^+$'s stopping in the same target. In this way two particles
of the two reactions cross approximately the same region of the
apparatus. The detector efficiencies roughly cancel out for tracks
of the same sign (muons and protons), and only the detector
efficiency for $\pi^-$'s and the global efficiencies of each
reaction must be evaluated, again by means of a dedicated
simulation of both the reaction under study and of the $K_{\mu2}$
decay occurring in the chosen target. 
%Again we assume that, if no
% selection on the direction is made, since the spatial region in the
%apparatus is spanned by almost similar tracks is the same
%independent of the particle charge, the detector efficiency for
%$\pi^-$ is roughly equal to that estimated for muons emitted
%from the same target.
Again we assume that, without any selection
on the direction, the detector efficiency for
$\pi^-$ is roughly equal to that estimated for muons emitted
from the same target. This is due to the fact that
almost similar tracks span is the same apparatus region,
independently of the particle charge.

The capture
rate $R'$ is therefore evaluated by the formula
\begin{equation}
R' = \frac{N(ev)}{N_{\mu^+}}\cdot \frac{N(K^+_{stop})}{N(K^-_{stop})}
\cdot \frac{\epsilon_G(\mu^+)}{\epsilon_G(ev)}
\cdot\frac{\epsilon_D(\mu^+)}{\epsilon_D(ev)}
\label{eq:rprime}
\end{equation}

Tab. \ref{tab:numeri2}  reports
the relevant numbers to get the $R'$ values
for the two $^6$Li targets.

\begin{table}[ht]
\centering
\begin{tabular}{lccc}
\hline
$N(ev)$ & $N_{\mu^+}$ & $N(K^+_{stop})$ & $N(K^-_{stop})$ \\ \hline
$\scriptstyle  113\pm 11_{stat}\pm 11_{sys}$
& $\scriptstyle 76475\pm 277$ 
& $\scriptstyle 1214996\pm 1102_{stat}\pm 60750_{sys}$ 
& $\scriptstyle 1535943\pm 1239_{stat}\pm 76797_{sys}$ \\
$\scriptstyle 132\pm 11_{stat}\pm 2_{sys}$ 
& $\scriptstyle 92031\pm 303$ & 
$\scriptstyle 1414073\pm 1189_{stat}\pm 70704_{sys}$
& $\scriptstyle 1771202\pm 1331_{stat}\pm 88560_{sys}$ \\
\hline
\end{tabular}
\begin{tabular}{lcc|c}
\hline
$\epsilon_G(\mu^+)$ &  $\epsilon_G(ev)\times 10^3$ & $\epsilon_D(\mu^+)$
& $R'/K^-_{stop}(\%)$ \\ \hline
$0.129\pm 0.002$ & $14.6\pm 0.5$ & $0.489\pm 0.011$ &
$2.22 \pm 0.45_{stat}\; ^{+0.22}_{-0.25}(sys)$ \\
$0.126\pm 0.002$ & $15.9\pm 0.4$ & $0.561\pm 0.011$ &
$1.50 \pm 0.27_{stat}\; ^{+0.15}_{-0.17}(sys)$ \\
\hline
\end{tabular}
\caption{Measured number of events, $K^-$ and $K^+$ stops in the same
$^6$Li target, and capture rate
 for the reaction $K^- + (pn)\rightarrow \Sigma^- p$ evaluated using
the method  of $\mu^+$ normalization ($K_{\mu 2}$ decays from
the same target as the reaction under study).
The number of selected events is already
background subtracted (the background amounts to about 30\% of the
observed signal).
The two lines correspond to events collected on two different $^6$Li targets.}
\label{tab:numeri2}
\end{table}

The values of $R$ and $R'$ for both the targets are in agreement,
within the statistical error, except for the value of $R'$ for the
first $^6$Li target, which is compatible with the $R$ measurement
from the same target within 2$\sigma$ only. We assume that a total
statistical error as big as the largest of the two measurements on
the same target can be assigned to their weighted mean value.

Therefore, one gets for the weighted means of the evaluated rates
for the two $^6$Li targets (denoted in the following as 1 and 2)
the two capture rates: $R_1 =
(1.94\pm 0.62_{stat}\; ^{+0.20}_{-0.22}(sys))\%/K^-_{stop}$ and
$R_2 = (1.57\pm 0.24_{stat}\;
^{+0.18}_{-0.37}(sys))\%/K^-_{stop}$; their mean value is
$(1.62\pm 0.23_{stat}\;^{+0.71}_{-0.44}(sys))\%/K^-_{stop}$ , to
be compared to previous measurements of the $K^-$ induced
reaction obtained in bubble chamber experiments. Ref.
\cite{bubble1} reports a multinucleonic $K^-$ capture rate at rest
of the order of $\sim 10\%$ in Deuterium, $\sim 16\%$ in Helium
and $\sim 20\%$ in Carbon: the absorption rate is assumed to
remain rather constant beyond $A=4$, perhaps increasing slowly for
heavier nuclei where the absorption most likely occurs on the
nuclear surface. Out of all the multinucleonic $K^-$ capture
events in Helium, according to Ref. \cite{bubble2} the rate for
$K^-\; ^4\mathrm{He}\rightarrow \Sigma^- pd$ is $(1.6\pm
0.6)\%/K^-_{stop}$, while for $K^-\; ^4\mathrm{He}\rightarrow
\Sigma^- ppn$ is $(2.0\pm 0.7)\%/K^-_{stop}$.

These  figures are in fair agreement with the results obtained in
the present analysis. They can be directly compared for the case
of the $K^-\; ^4\mathrm{He}$ reaction where the $d$ or $(np)$ in
the final state can be assumed as spectators.

\section{Conclusions}
The inclusive proton momentum spectrum in $K^-$ reactions induced
in $^6$Li exhibits a sharp structure at about 500 MeV/$c$ which
can be rather easily explained as due to the interaction of the
$K^-$ on a ``quasi"-deuteron cluster in the $^6$Li nucleus,
leading to the two-body final state $\Sigma^- + p$. Both the
observed proton and the $\pi^-$ from the $\Sigma^-$ decay have the
right kinematic features to confirm that they belong to this
reaction. Moreover, the rate for such a reaction is compatible
with the rates of similar reactions observed in helium bubble
chamber experiments.

Since the signal is quite similar to the observations of E471 in
$K^-$ induced reactions on a $^4$He target, 
the peaks observed by the two experiments could have the same
nature, and maybe they are simply due to a two-body reaction where
a $K^-$ interacts on a ``quasi" deuteron, both in $^6$Li and in
$^4$He. This possibility makes the claim for a deeply bound state
unnecessary. On the other hand, the absence of such a signal in
the spectra out of $^{12}$C targets strengthens this hypothesis,
as a similar two-body mechanism would be possible even on $^{12}$C
nuclei, but very difficult to observe because of the size of the whole
nuclear system and the dilution effects it entails.

Thus, we point out that the possibility of selective capture by
clusters in nuclei can be the source of ambiguities in missing
mass experiments. Invariant mass measurements, exclusive and more
precise, would probably help to better clarify the interesting
subject of the existence of such states, supported by complementary 
information obtained with the missing mass method.
In fact, it must be reminded that the
first experimental observation with the invariant mass technique,
performed by FINUDA, although more precise, 
presents a few inconsistencies compared to the expectations of some
theoretical models. So, on one side more experimental
data are urgently needed, to confirm the observed structures and to check
for possible different decay patterns. 
On the other hand,
further theoretical calculations are required to explain the
experimental observations and possibly discard their interpretation 
as $\overline K$-nuclear states. 

\section{Acknowledgments}
We are indebted to Prof. T. Yamazaki for enlightening
discussions and Prof. A. Gal for precious suggestions and remarks.
We thank the DA$\Phi$NE crews for their skillful handling of the
collider and the FINUDA technical staff for their constant
support.


\begin{thebibliography}{99}
\bibitem{akaishi} Y.~Akaishi and T.~Yamazaki, {\it Phys. Rev.} {\bf C65}  (2002), 044005
\bibitem{dote} A. Dot\'e {\it et al.}, {\it Phys. Lett.} {\bf B590} (2004), 51;\\
A. Dot\'e {\it et al.}, {\it Phys. Rev. C} {\bf 70} (2004), 044313
\bibitem{ref2} T.~Suzuki {\it et al.}, {\it Phys. Lett.} {\bf B597} (2004), 263
\bibitem{ref3} T.~Suzuki {\it et al.}, {\it Nucl. Phys.} {\bf A754} (2005), 375c
\bibitem{ref4} M.~Agnello {\it et al.}, {\it Phys. Rev. Lett.} {\bf 94} (2005), 212303
\bibitem{kishimoto} T.~Kishimoto {\it et al.}, {\it Nucl. Phys.} {\bf A754} (2005), 383c
\bibitem{re:shallow}J. Sch\"affner-Bielich {\it et al}, {\it Nucl. Phys.}
{\bf A669} (2000), 153;\\
A. Ramos and E. Oset, {\it Nucl. Phys.} {\bf A671} (2000), 481; \\
A. Cieply {\it et al.}, {\it Nucl. Phys.} {\bf 696} (2001), 173
\bibitem{galultimo} J. Mare$\check{\mathrm s}$ {\it et al.},
{\it Nucl. Phys.} {\bf A770} (2006), 84
\bibitem{osetprep} E. Oset and H. Toki, preprint nucl-th/0509048 \\
V.K. Magas {\it et al.}, preprint nucl-th/0601013
\bibitem{re:hadron} A. Filippi {\it et al.}, FINUDA Collaboration, in
{\it Proceedings of Eleventh International Conference on Hadron
Spectroscopy,}, Rio de Janeiro, Brazil, 21--26 August 2005, ed. A.
Reis {\it et al.}, AIP Conference Proceedings {\bf 814}, Melville,
New York (2006), p. 598
\bibitem{ref6} M.~Agnello {\it et al.}, {\it Nucl. Phys.} {\bf A754} (2005), 399c
\bibitem{ref7} M.~Agnello {\it et al.}, {\it Phys. Lett.} {\bf B622} (2005), 35
\bibitem{ref8} A.~Zenoni in {\it Proc. Int. School of Physics ``E. Fermi''} Course CLVIII
(Hadron Physics), eds T. Bressani, U. Wiedner and A. Filippi,
S.I.F. - IOS Press (2005), 183
\bibitem{ref14} T.~Bressani {\it et al.}, {\it Nucl. Phys.} {\bf B9} (1969), 427
\bibitem{ref15} J.~Favier {\it et al.}, {\it Nucl. Phys.} {\bf A169} (1971), 540
\bibitem{ref16} C.~Cernigoi {\it et al.}, {\it Nucl. Phys.} {\bf A352} (1981), 343
\bibitem{ref17} C.~Cernigoi {\it et al.}, {\it Nucl. Phys.} {\bf A456} (1986), 599
\bibitem{yamahiren} T. Yamazaki and S. Hirenzaki, {\it Phys. Lett.} {\bf B557}
(2003), 20 
\bibitem{geane} V. Innocente {\it et al.}, GEANE, Average Tracking and Error
Propagation Package, CERN Program Library, W5013-E, 1991
\bibitem{ref11} J.~Mare$\check{\mathrm s}$ {\it et al.}, {\it Phys. Lett.} {\bf B606} (2005), 295
\bibitem{bubble1} C. Vander Velde-Wilcquet {\it et al.}, {\it Il Nuovo Cimento} {\bf 39A} (1977), 538
\bibitem{bubble2} P.A. Katz {\it et al.}, {\it Phys. Rev.} {\bf D1}
(1970), 1267

\end{thebibliography}
\end{document}